\newcommand{\ud}{\rm d}
\newcommand{\un}{~\mathrm}
\newcommand{\ie}{{\em i.e.} }
\newcommand{\eg}{{\em e.g.} }
\documentclass[twocolumn,showpacs,preprintnumbers,amsmath,amssymb,superscriptaddress]{revtex4}

\usepackage{graphicx}
\usepackage{dcolumn}
\usepackage{bm}
\usepackage{amsmath}

\begin{document}

\title{Numerical simulation of 2D steady granular flows in rotating drum: On surface flows rheology}
\author{M. Renouf}
\affiliation{Projet SIAMES/BIPOP, IRISA / INRIA Rennes, Campus Universitaire de Beaulieu, Avenue du G{\'e}n{\'e}ral Leclerc, 35042 RENNES Cedex - France}
\author{D. Bonamy}
\affiliation{Groupe fracture, DSM/DRECAM/SPCSI, CEA Saclay, F-91191 Gif sur Yvette - France}
\author{F. Dubois}
\affiliation{Laboratoire de M{\'e}canique \& G{\'e}nie Civil  - UMR CNRS-UM2 5508, Universit{\'e} Montpellier 2, C.C. 48 - Place Bataillon, F-34095 Montpellier Cedex 5 - France}
\author{P. Alart}
\affiliation{Laboratoire de M{\'e}canique \& G{\'e}nie Civil  - UMR CNRS-UM2 5508, Universit{\'e} Montpellier 2, C.C. 48 - Place Bataillon, F-34095 Montpellier Cedex 5 - France}

\begin{abstract}
The rheology of 2D steady surface flow of cohesionless cylinders in a rotating drum is investigated through {\em Non Smooth Contact Dynamics} simulations. Profile of volume fraction, translational and angular velocity, rms velocity, strain rate and stress tensor were measured at the midpoint along the length of the surface flowing layer where the flow is generally considered as steady and homogeneous. Analysis of these data and their inter-relations suggest the local inertial number - defined as the ratio between local inertial forces and local confinement forces - to be the relevant dimensionless parameter to describe the transition from the quasi-static part of the packing to the flowing part at the surface of the heap. Variations of the components of the stress tensor as well as the ones of rms velocity as a function of the inertial number are analysed within both the quasi-static and the flowing phases. Their implications are discussed.

\end{abstract}

\pacs{45.70.-n, 
            83.70.Fn,  
            46.10.+z   
}

\date{\today}
\maketitle


\section{Introduction}\label{intro}

Granular media present numbers of interesting and unusual behaviours: They can flow as liquids, but, under some circumstances, they can jam and resist to external shear stress without deforming. Understanding rheology of granular systems has thus developed along two major themes: The {\em Rapid flows} - gaseous-like - regime where grains interact through binary collisions, are generally described in the framework of the kinetic theory~\cite{Savage81_jfm,Jenkins83_jfm,Lun86_am}; The {\em slow flow} - solid-like - regime where grain inertia is negligible is most commonly described using the tools of soil mechanics and plasticity theory \cite{nedderman92_book}. In between these two regimes there exists a {\em dense flow} - liquid-like - regime where grain inertia becomes important but contacts between grains are kept. Rheology of this last regime has been widely investigated experimentally, numerically and theoretically (see \cite{Gdrmidi04_epje} for a review), but still remains far from being understood. Several models have been proposed recently to describe dense granular flows by accounting for non-local effects~\cite{Mills99_epl,Andreotti01_pre,Jenkins02_pof,Bonamy03_epl,Rajchenbach03_prl}, by adapting kinetic theory~\cite{Savage98_jfm,Bocquet02_pre,Mohan02_jfm}, by modelling dense flows as partially fluidized flows~\cite{Aranson02_pre,Josserand04_epje} or by considering them as quasi-static flows where the mean motion results from transient fractures modelled as self activated process~\cite{Pouliquen96_pre,Debregeas00_epl,Pouliquen01_acs,Lemaitre02_prl}, but, to our knowledge, none of them succeed to account for all the features experimentally observed.

The most spectacular manifestation of this solid/liquid duality occurs during an avalanche when a thin layer of grains starts to roll at the surface of the packing, most of the grains remaining apparently static. The global evolution of such surface flows can be captured by models derived from non-linear physics~\cite{Bouchaud94_jpi, Boutreux98_pre,Aradian98_pre} or fluid mechanics~\cite{savage89_jfm,Douady99_epjb,Khakhar97_pof,Bonamy02_pof}. However, some experimental results remain unexplained: For instance, experimental velocity profiles measured in two-dimensional (2D) flows~\cite{Rajchenbach00_ap,Bonamy01_phd,Bonamy03_gm} or three dimensional (3D) flows~\cite{Bonamy02_pof,Orpe01_pre,Jain02_pof,Felix02_phd} clearly exhibit the selection of a constant velocity gradient within the flowing layer while momentum balance implies that the shear stress increases linearly with depth. This observation is incompatible with any local and one-to-one stress/strain constitutive relations. Recent experiments~\cite{Bonamy02_prl} have provided evidence of 'jammed' aggregates embedded in the avalanche. These 'solid' clusters are found to be power-law distributed without any characteristic length-scales, and may explain the failure of present models. But a clear understanding of the avalanche rheology is still missing.

The purpose of this paper is to investigate the surface flows rheology through numerical simulations of 2D 'minimal' granular systems made of cohesionless weakly polydisperse cylinders confined in a slowly rotating drum. Those allow us to track the evolution of quantities like stress that are not accessible in real experiments. Moreover, they allow to get rid of artefacts such as the friction of beads on the lateral boundaries of the drum that may confuse the interpretation of an experiment. The numerical simulation were performed using contact dynamic methods~\cite{Jean92_conf,Moreau94_ejmas} based on a fully implicit resolution of the contact forces, without any resort to regularization schemes. At a given step of evolution, all the kinematic constraints within the packing are simultaneously taken into account together with the equations of motions to determine all the contact forces in the packing. This allows to deal properly with nonlocal momentum transfers implied in multiple collisions, contrary to Molecular Dynamics schemes traditionally used that reduce the system evolution to a succession of binary collisions.

The simulation scheme and the description of the simulated systems are detailed in Sec. II. In Sec. III, we report comprehensive analysis of volume fraction and velocity (translational and rotational) profiles at the center of the drum. They are compared with experimental data available in the literature. Stress analysis and implications on the rheology of free surface flows are discussed in Sec. IV. and Section V respectively. Sec. VI is focussed on the analysis of both the translational and rotational velocity fluctuations. Finally, Sec. VII summarizes our findings.

\section{Simulation Methodology}\label{setup}

For this study, we simulate granular systems similar to those investigated experimentally in~\cite{Rajchenbach00_ap,Bonamy01_phd,Bonamy02_prl,Bonamy03_gm}. We model a 2D rotating drum of diameter $D_0$ equal to $450\un{mm}$ half-filled with $7183$ rigid disks of density $\rho_0=2.7\un{g.cm}^{-3}$ and diameter uniformly distributed between $3\un{mm}$ and $3.6\un{mm}$. This weak polydispersity prevents any 2D ordering effect that may induce nongeneric effects. Normal restitution coefficient between two disks (respectively between disks and drum) is set to $0.46$ (respectively $0.46$) and the friction coefficient to $0.4$ (respectively $0.95$). Normal restitution coefficients and disk/disk friction coefficient were chosen to mimic the experimental flows of aluminium beads investigated in \cite{Bonamy01_phd,Bonamy03_gm}. The drum/disk friction coefficient was set close to $1$ to prevent sliding at the drum boundary.

Numerical simulations dedicated to evolution of granular media can be based either on explicit~\cite{Drake91_jfm,Cundall79_g,Kishino88_conf,Silbert01_pre} or implicit~\cite{Jean92_conf,Moreau94_ejmas,DeSaxe91_jmsm} method. One of the drawbacks of explicit models is to reduce non-local momentum transfers implied in multiple collisions to a succession of binary collisions. Moreover, numerical instabilities can occur in granular flows. They are corrected either by introducing some artificial viscosity or by reducing the size of the time step. The \textit{Non Smooth Contact Dynamics} method used here is implicit. It provides a nonsmooth formulation of the body's impenetrability condition, the collision rules and the dry Coulomb friction law. The method is extensively described in~\cite{Jean99_cmame}, and briefly explained below.

Firstly, equations of motion are written for a collection of rigid bodies and discretized by a time integrator \cite{Moreau99_cmame}. The interaction problem is then solved at contact level (local level) rather than at particle level (global level) as commonly performed in explicit methods. In other words, equations are written in term of relative velocities $\mathbf{u}_\alpha$ and local impulsions $\mathbf{r}_{\alpha}$ defined at each contact point indexed by $\alpha$. The impenetrability condition evoked previously means that particles candidates for contact should not cross the boundaries of antagonist's bodies. We consider also that contacting bodies do not attract each other, \ie that the reaction force is positive, and vanishes when the contact vanishes. This can be summarized in the following so-called \textit{velocity Signorini Condition}:
\begin{equation}
u_{n} \ge 0 \quad r_{n} \ge 0 \quad u_{n}.r_{n}=0,
\end{equation}
where the index $n$ denotes the normal component of the various quantities (index $\alpha$ is omitted). Let us note that this philosophy is different from what is used in explicit methods, where normal forces are usually proportional to the penetration between two particles. The dry frictional law is the Coulomb's one for which the basic features are: The friction force lies in the Coulomb's cone ($||r_{t}|| \le \mu r_n$, $\mu$ friction coefficient), and if the sliding relative velocity is not equal to zero, its direction is opposed to the friction force ($||r_{t}|| = \mu r_n$). This summarized in the following relation:
\begin{equation}
||r_{t}|| \le \mu r_n \quad ||u_t|| \ne 0 \rightarrow r_t = -\mu r_n \frac{u_t}{||u_t||}
\end{equation}
For rigid bodies we also need to adopt a collision law because the velocity Signorini condition does not give enough information. We adopt the Newton restitution law, $u_{n}^+=-e_n u_{n}^-$, realistic for collection of disks. The reader can refer to \cite{Moreau88_conf} for more explanations about collision laws. Time discretization of equations of motion - where the global contact forces are the only missing quantities to determine the motion of each bead - lead to the following scheme:
\begin{equation}
\label{eq:1}
\left\{ \begin{array}{l}
\mathbb{W}\mathbf{r} - \mathbf{u} =\mathbf{b}\\
law_\alpha[\mathbf{u}_\alpha,\mathbf{r}_\alpha]=.true.,\quad \alpha=1,n_c
\end{array} \right.
\end{equation}
where $\mathbf{u}$ and $\mathbf{r}$ denotes the vectors containing the relative velocity and the mean contact impulse for {\em all} the contact points respectively. The matrix $\mathbb{W}$ is the Delassus operator \cite{Delassus17_asens} that contains all the local informations (local frames and contact points) as well as the informations related to the contacts connectivity. The right hand side of first line in Eq.~\ref{eq:1} represents the free relative velocity calculated by only taking into account the external forces. The operator $law_\alpha$ encodes the friction-contact law which should be satisfied by each component of couple $(\mathbf{u}_\alpha,\mathbf{r}_\alpha)$; $n_c$ denotes the number of contact. Systems of Eq.~\ref{eq:1} can be solved by a classical non linear Gau$\beta$-Seidel algorithm \cite{Jean99_cmame} or a Conjugate Projected Gradient one \cite{Renouf04_came}. This two algorithms benefit from parallel versions \cite{Renouf04_jcam,Renouf04_reef} which show their efficiency in the simulations of large systems. Information from this local level, the contact level, is transfered to the global level, the grain level and the configuration of the system is updated.

The procedure to achieve a numerical experiment is the following: All the disks are placed in an immobile drum; Once the packing is stabilized, a constant rotation speed $\Omega$ (ranging from $2\un{rpm}$ to $15\un{rpm}$) is imposed to the drum; After one round, a steady continuous surface flow is reached (This has been checked by looking at the time evolution of the total kinetic energy within the packing over the next round); One starts then to capture 400 snapshots equally distributed over a rotation of the drum. The time-step is set to $6.10^{-3} \un{s}$. The number of time-steps necessary to achieve an experiment ranges from $4.10^{3}$ to $10^{4}$ depending on the rotating speed. All simulations have been performed with LMGC90 software~\cite{Dubois03_conf}. On a SGI Origin 3800 with 16 processors, about $20\un{h}$ are required to achieve one of these simulations.

\begin{figure}[!tpb]
\centering
\includegraphics[width=0.8\columnwidth]{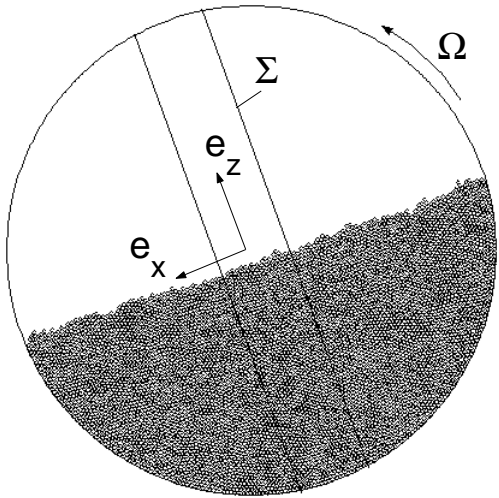}
\caption{Typical snapshot of the steady surface flows in the simulated 2D rotating drum. Its diameter and its rotation speed are respectively $D_0=450\un{mm}$ and $\Omega=6\un{rpm}$. It is filled with $7183$ disks of density $\rho_0=2.7\un{g.cm}^{-3}$ and diameter uniformly distributed in the interval $[3\un{mm};3.6\un{mm}]$. the disk/disk coefficient of restitution and disk/disk friction coefficient are set respectively to $e_n = 0.46$ and $\mu=0.2$. the disk/drum coefficient of restitution and disk/drum friction coefficient are set respectively to $e_n^0=0.46$ and $\mu^0=0.95$.}
\label{snapshot}
\end{figure}

A typical snapshot of the simulated granular packing is shown in Fig.~\ref{snapshot}. For each bead of each of the $400$ frames within a given numerical experiment, one records the position $\mathbf{x}$ of its center of mass, the "instantaneous" velocity $\mathbf{c}$ of this center of mass measured over a time window $\delta_t=6.10^{-3}\un{s}$ and its angular velocity $w$. Vorono{\"i} tessellation was used to define the local volume fraction associated to each bead (see {\em e.g.}~\cite{Bonamy02_prl}). The components of contact stress tensor $\mathbf{\sigma}$ associated to each bead $i$ are computed as~\cite{Radjai96_prl,Radjai98_prl}:

\begin{equation}
\sigma_{\alpha\beta}=\frac{1}{2V_i}\sum_{j\not{=}i} x_{ji}^\alpha F_{ji}^\beta, \quad \alpha,\beta \in \{1,2\},
\end{equation}

\noindent where $V_i$ is the volume of the Vorono{\"i} polyedra associated to the bead $i$, ${\mathbf F}_{ji}$ the contact force between $i$ and $j$, and $\mathbf{x}_{ji}=\mathbf{x}_j-\mathbf{x}_i$. In all the following, these quantities are nondimensionalized: Calling $g$ the gravity constant and $d$ the mean disk diameter, distances, time, velocities and stresses are given in units of $d$, $\sqrt{d/g}$, $\sqrt{gd}$ and $\rho_0 g d$ respectively. In this paper, we concentrate on the continuum scale by looking at profiles of the time and space averaged quantities. Statistical analysis of these quantities at the grain scale will be presented in a separated paper.

In rotating drum geometries, the surface flow is not fully developed. The frame of study should now be chosen appropriately. One thus define the frame $\Re$ {\em rotating with the drum} that coincides with the reference frame $\Re_0=(\mathbf{e}_x,\mathbf{e}_z)$ fixed in the laboratory, so that  $\mathbf{e}_x$ (resp. $\mathbf{e}_z$) is parallel (resp. perpendicular) to the free surface (see Fig.~\ref{snapshot}) \cite{defAngle}. In the frame $\Re$, the flow can be considered as quasi-homogeneous at the center of the drum, \eg within the elementary slice $\Sigma$ (see Fig.~\ref{snapshot}) 20 beads diameter wide, parallel to $\mathbf{e}_z$ located at $x=0$. This slice is divided into layers of one mean bead diameter wide parallel to the flow. The given value of a given continuum quantity $\bar{a}(z)$ (volume fraction, velocity, stress...) at depth $z$ is then defined as the average of the corresponding quantity defined at the grain scale over all the beads in all the 400 frames of the sequence whose center of mass is inside the layer.

\section{Kinematic analysis}\label{kinematic}

\subsection{Volume fraction profile}

\begin{figure}[!tpb]
\centering
\includegraphics[height=0.6\columnwidth]{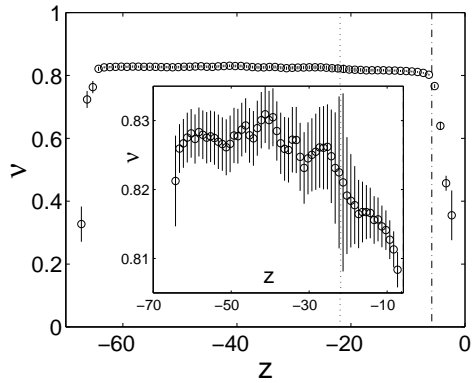}
\caption{Volume fraction profile $\nu(z)$ (averaged over 400 frames) at the center of the drum obtained for $\Omega=6\un{rpm}$. The errorbars correspond to a $99\%$ confident interval. The vertical mixed line locates the free surface boundary, -~set to the point where $\nu$ crosses the value $0.7$~-, from which  $\nu$ drops quickly down to zero. The vertical dotted line locates the static/flowing interface, defined from the streamwise velocity profile (see Fig.~\ref{vit6rpm}). Apart from these two zones, $\nu$ appears almost constant, around $0.8$. Inset, zoom in the "constant region" enhancing the small variations of $\nu$ within the two phases.}
\label{comp6rpm}
\end{figure}

Let us first focus on volume fraction profiles within the packing. Figure~\ref{comp6rpm} displays the volume fraction profile measured for $\Omega=6\un{rpm}$. To check the homogeneity of the flow with regard to $\nu$, the elementary slice $\Sigma$ was translated of an increment of 5 bead diameters in both positive $x$ and negative $x$. The volume fraction profile is found to be invariant under infinitesimal translation along $\vec{e}_x$. At the free surface, $\nu$ drops quickly to zero within a small zone of thickness around three/four beads diameter independent of $\Omega$. In all the following, the free surface boundary is set at the lower boundary of this small region (mixed line in Fig.~\ref{comp6rpm}), defined at the point where $\nu$ becomes larger than $0.7$. At the drum boundary, $\nu$ jumps also to a much smaller value within a small region about two/three beads diameters thick, which should be attributed to the presence of the smooth drum boundary. Apart fr21.8om these two narrow regions, the volume fraction $\nu$ is almost constant within the drum, around the random close packing (RCP) value $\nu^{RCP} \simeq 0.82$. A closer look (Inset of Fig.~\ref{comp6rpm}) suggests that $\nu$ is constant within the static phase, and decreases weakly within the flowing layer as defined from the velocity profile in next section. Such behaviour is expected since granular systems should dilate before being able to deform. However, this decreasing is very small and compressible effects can thus be neglected with regard to momentum balance, even if they may significantly alter the local flow rheology~\cite{Bonamy02_prl}.

\subsection{Velocity profiles}

\begin{figure}[!tpb]
\centering
\includegraphics[height=0.6\columnwidth]{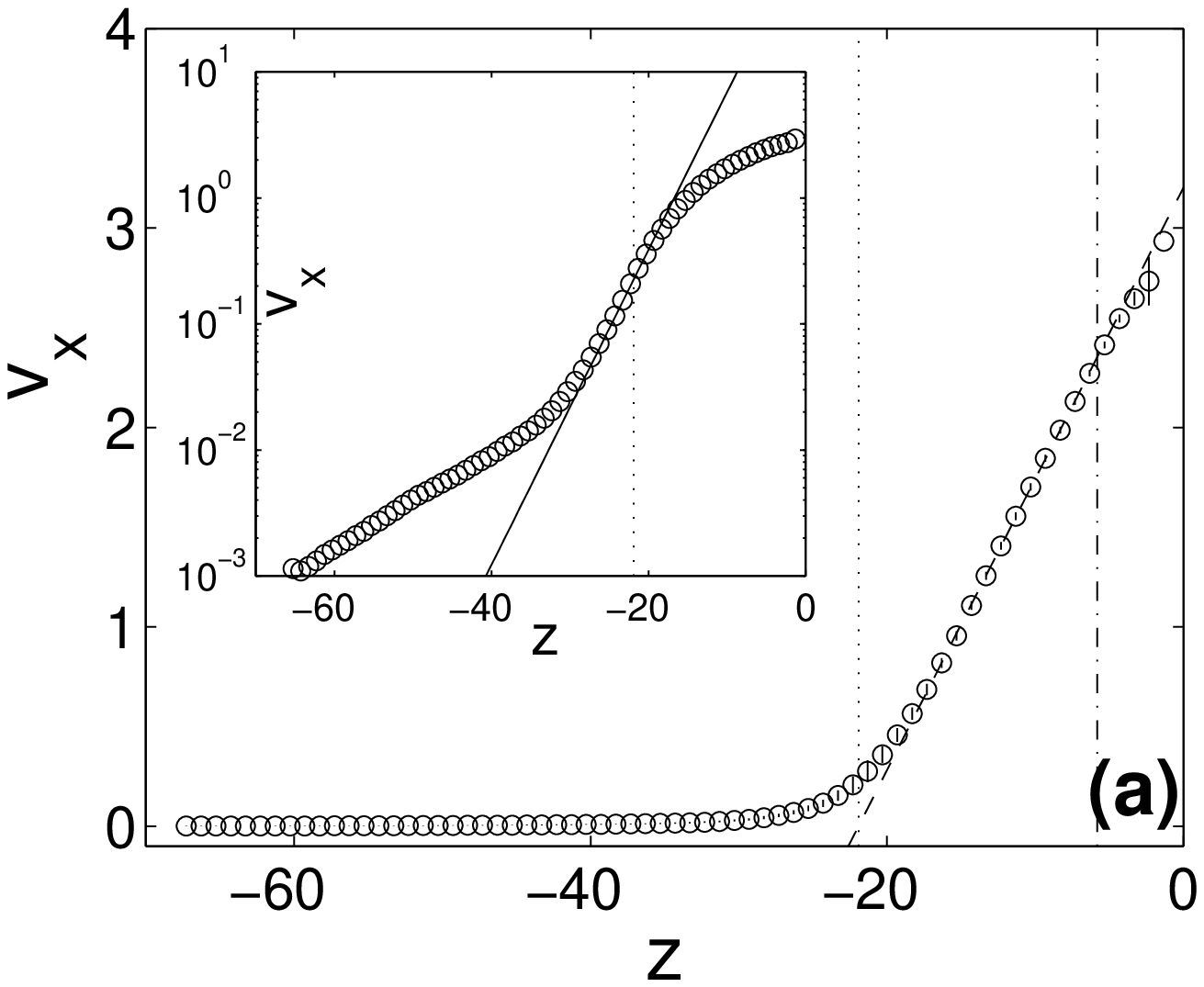}
\includegraphics[height=0.6\columnwidth]{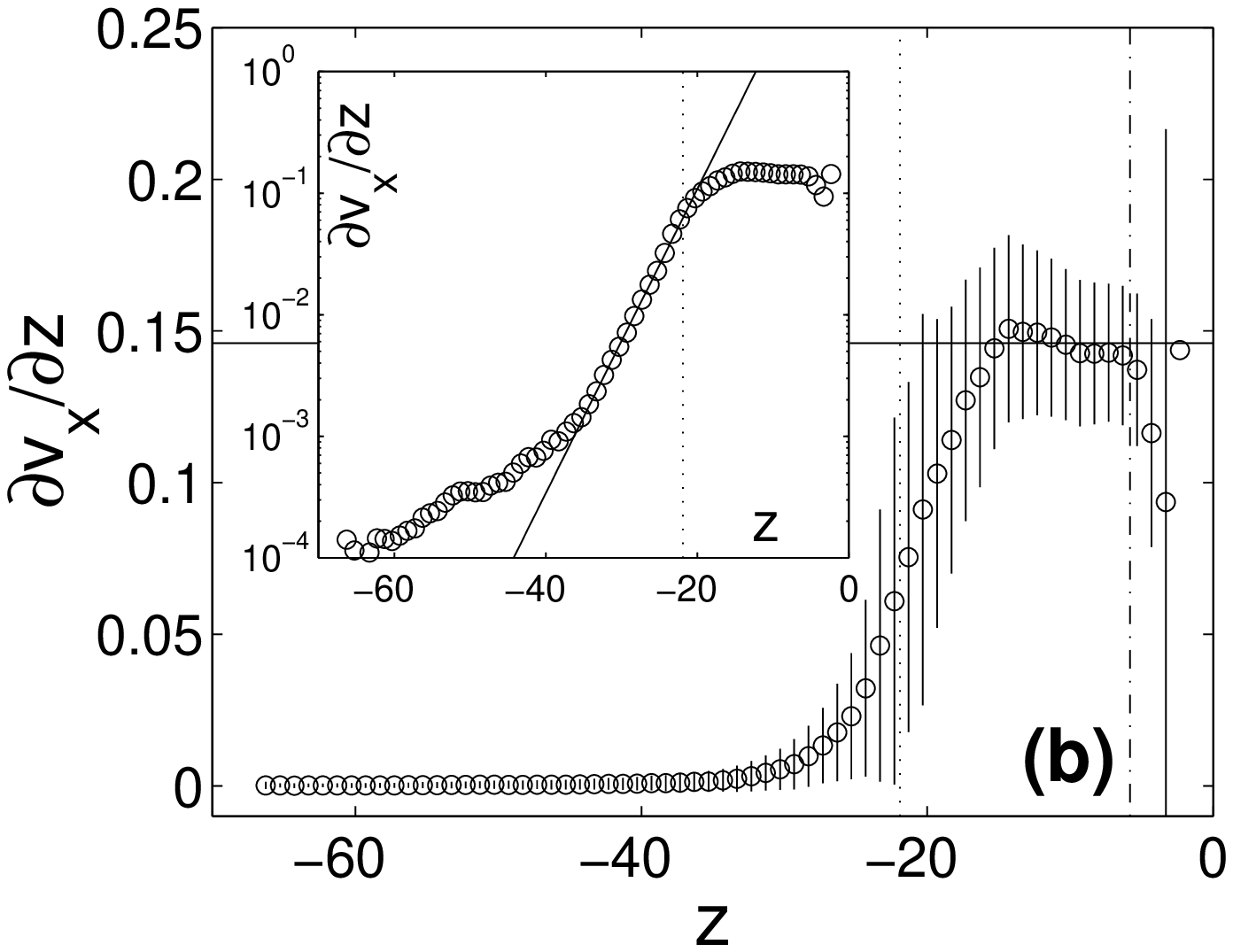}
\caption{Profile of (a) streamwise velocity $v_x(z)$ and (b) streamwise velocity gradient $\partial_z v_x(z)$ (averaged over 400 frames) at the center of the drum as obtained for $\Omega=6\un{rpm}$. The errorbars correspond to a $99\%$ confident interval. The profile of the velocity gradient (resp. of the velocity gradient) is linear (resp. constant) within the flowing layer. The plain straight line in sub-figure (a) (resp. in sub-figure (b)) corresponds to a linear fit: $v_x=\dot{\gamma}(z+H)$ (resp. a constant $\partial_z v_x=\dot{\gamma}$) where $\dot{\gamma} \simeq 0.15$. The vertical mixed line locates the free surface boundary (see Fig.~\ref{comp6rpm}). The flowing/static interface (dotted line) is defined from the depth where the straight line intersects the $z$-axis in sub-figure (a). The flowing layer thickness $H$ can then be deduced: $H = 14.6$. Inset of sub-figure (a) (resp. inset of sub-figure (b)): plot of the profile of the velocity $v_x(z)$ (resp. velocity gradient $\partial_z v_x(z)$) in semilogarithmic scales. In both insets, the plain straight line corresponds to an exponential fit of caracteristic decay length $\lambda \simeq 3.4$.}
\label{vit6rpm}
\end{figure}

As expected for a quasi-homogeneous flow, the normal component $v_z$ of the velocity was found to be negligible compared to the tangential component $v_x$ at any depth $z$. Figure \ref{vit6rpm} depicts both the streamwise velocity profile $v_x(z)$ (Fig. \ref{vit6rpm}a) and the shear rate profile $\partial_z v_x(z)$ (Fig. \ref{vit6rpm}b) for $\Omega=6\un{rpm}$. Both these profiles were found to be invariant under infinitesimal translation along $\vec{e}_x$. Two phases can clearly be observed: A flowing layer exhibiting a linear velocity profile and a static phase experiencing creep motion where both $v_x$, and $\partial v_x/\partial z$ decay exponentially with depth (see Inset of Fig.~\ref{vit6rpm}b). Such shapes are very similar to the ones observed experimentally in 2D flows~\cite{Rajchenbach00_ap,Bonamy01_phd,Bonamy03_gm} as well as in 3D flows~\cite{{Bonamy02_pof,Orpe01_pre,Jain02_pof,Felix02_phd}}. The interface between the two phases can then be defined by extrapolating the linear velocity profile of the flowing phase to zero (see Fig.~\ref{vit6rpm}a). The flowing layer thickness $H$ can then deduced.

\begin{figure}[!htb]
\centering
\includegraphics[height=0.6\columnwidth]{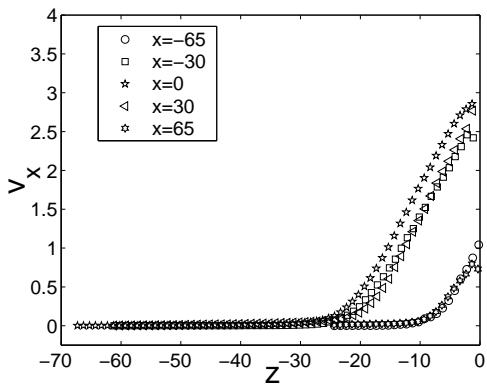}
\caption{Velocity profiles obtained for $\Omega=6\un{rpm}$ at five different locations $x$. The velocity gradient within the flowing layer as well as the exponential decreasing within the static phase are found to depend weakly on the precise location in the drum. They are consequently independent of both the flowing layer thickness $H$ and its $x$ derivative $\partial_x H$}
\label{velocityvsx}
\end{figure}

Velocity profiles measured for $\Omega=6\un{rpm}$ at five different locations $x$ are represented in Fig.~\ref{velocityvsx}. At these locations, $\partial_x H$ is no more equal to zero. The width of the elementary slice $\Sigma$ has thus been decreased to two beads diameters in order to minimize this drift. The shape of the velocity profile remains the same in these locations, with a clear linear velocity profile within the flowing layer and an exponentially decaying velocity within the static phase.  Both the characteristic decay length $\lambda$ of the exponential creep within the static phase and the constant velocity gradient $\dot{\gamma}_0$ within the flowing layer are observed to be independent of the precise location $x$ for a given value of $\Omega$.

The "natural" control parameter in our experiment is the rotating speed $\Omega$. However, comparisons between experiments in heap geometry and rotating drum geometry \cite{Gdrmidi04_epje} suggest that the main control parameter for the surface flow is the non-dimensionalized flow rate $Q$, defined as:

\begin{equation}
Q=\int_{z=-R_0}^{0}\nu(z)v_x(z)\ud z
\end{equation}

\noindent Its variation as well as the one of the flowing layer thickness $H$ and the mean slope $\theta$ with respect to $\Omega$ are reported in Tab.\ref{tabQvsO}.

\begin{table}[!hhh]
\begin{tabular*}{0.78\columnwidth}{|c|cccccc|}
\hline
$\Omega$ & 2 rpm & 4 rpm & 5 rpm & 6 rpm & 10 rpm & 15 rpm \\
\hline
$Q$ & 8 & 15.3 & 19.5 & 21.8 & 39.8 & 57.7 \\
$\theta$ & 18.1 & 18.9 & 19.3 & 19.7 & 21.2 & 23.0\\
$H$ & 9 & 11.5 & 13.3 & 14.6 & 17.2 & 21.0\\
\hline
\end{tabular*}
\caption{Variation of the nondimensionalized flow rate $Q$, the mean angle $\theta$ of the flow and the flowing layer thickness $H$ with respect to the rotating speed $\Omega$ within the elementary slice $\Sigma$ at the center of the drum.}
\label{tabQvsO}
\end{table}

\begin{figure}[!htb]
\centering
\includegraphics[height=0.6\columnwidth]{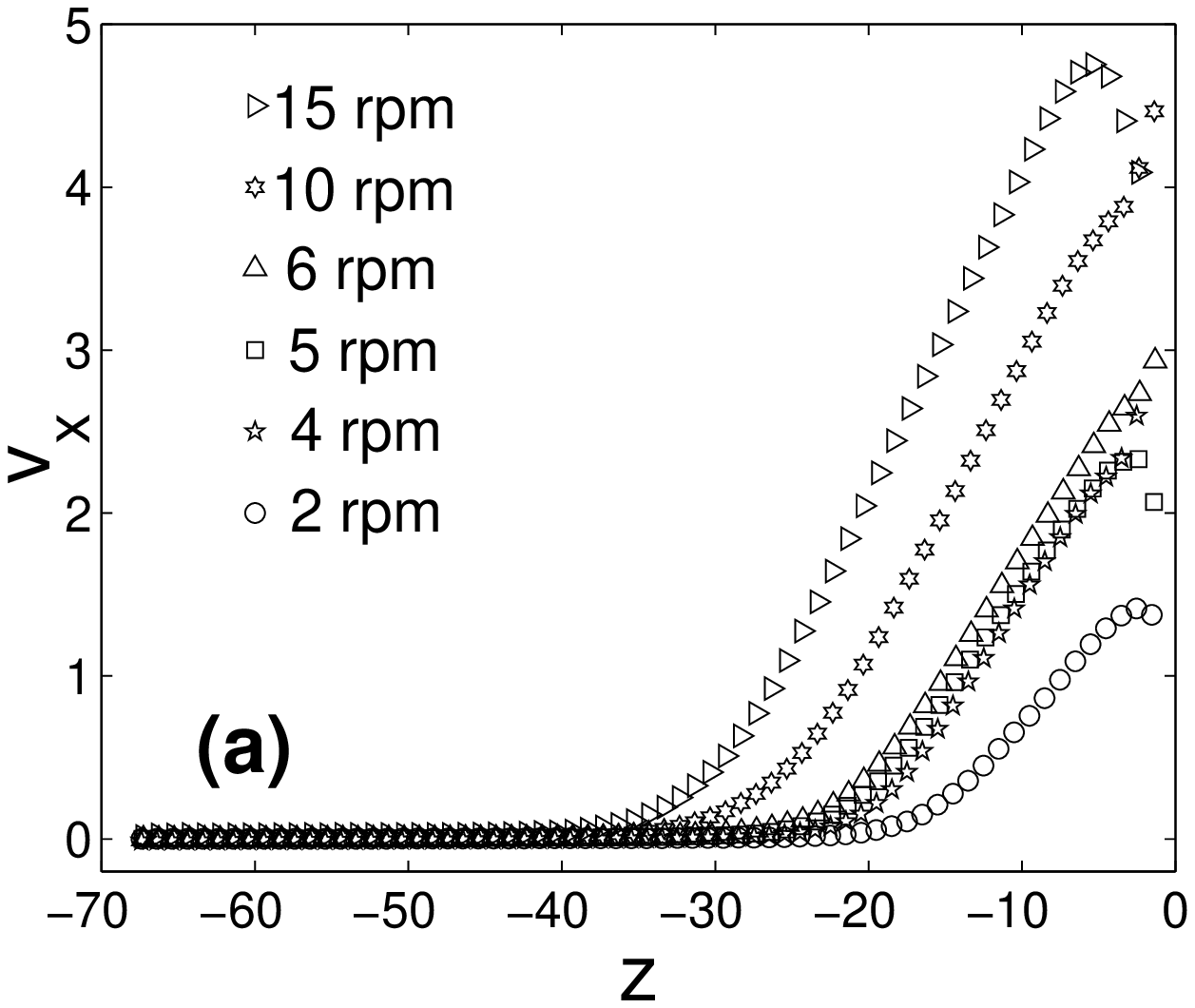}
\includegraphics[height=0.6\columnwidth]{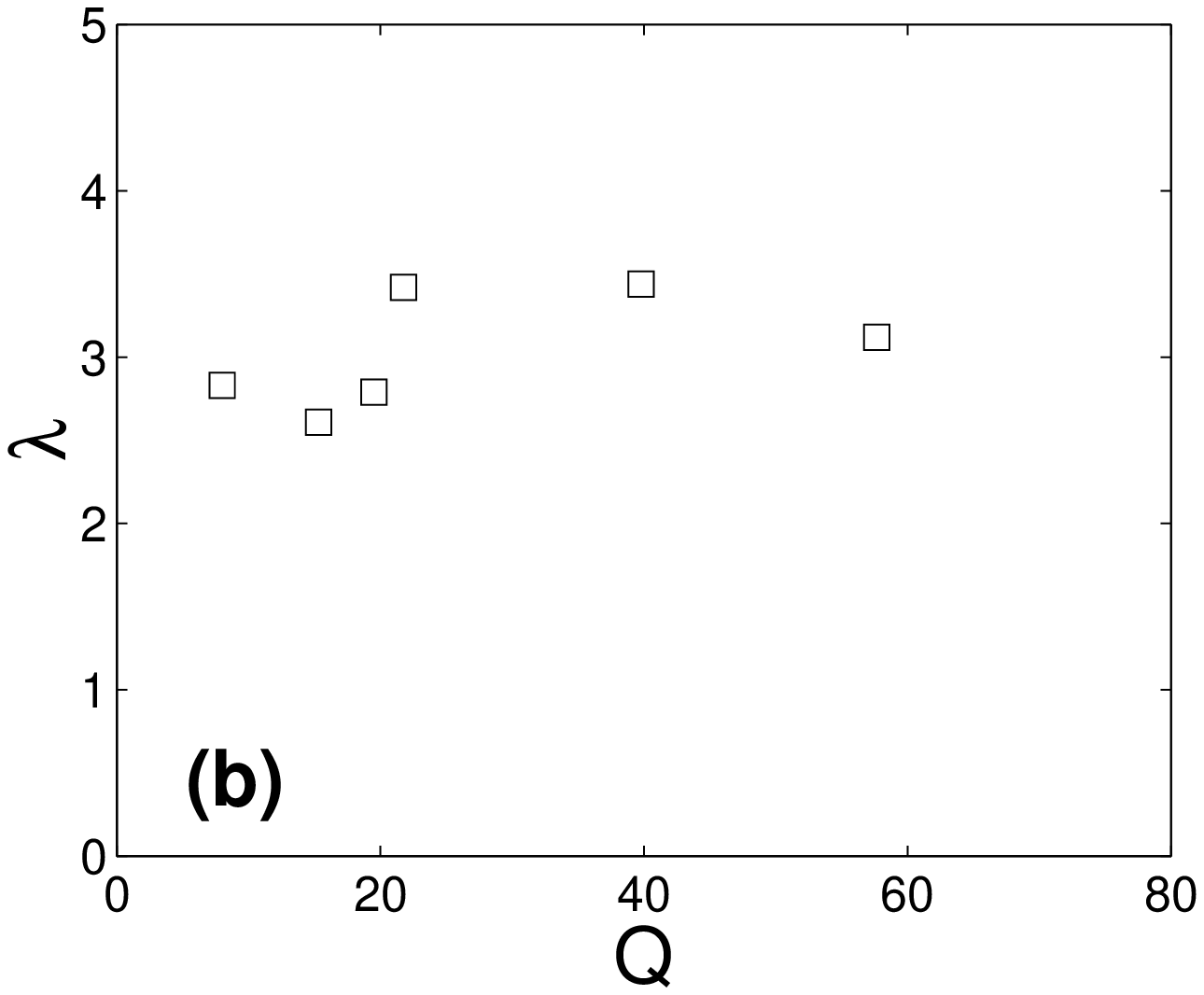}
\includegraphics[height=0.6\columnwidth]{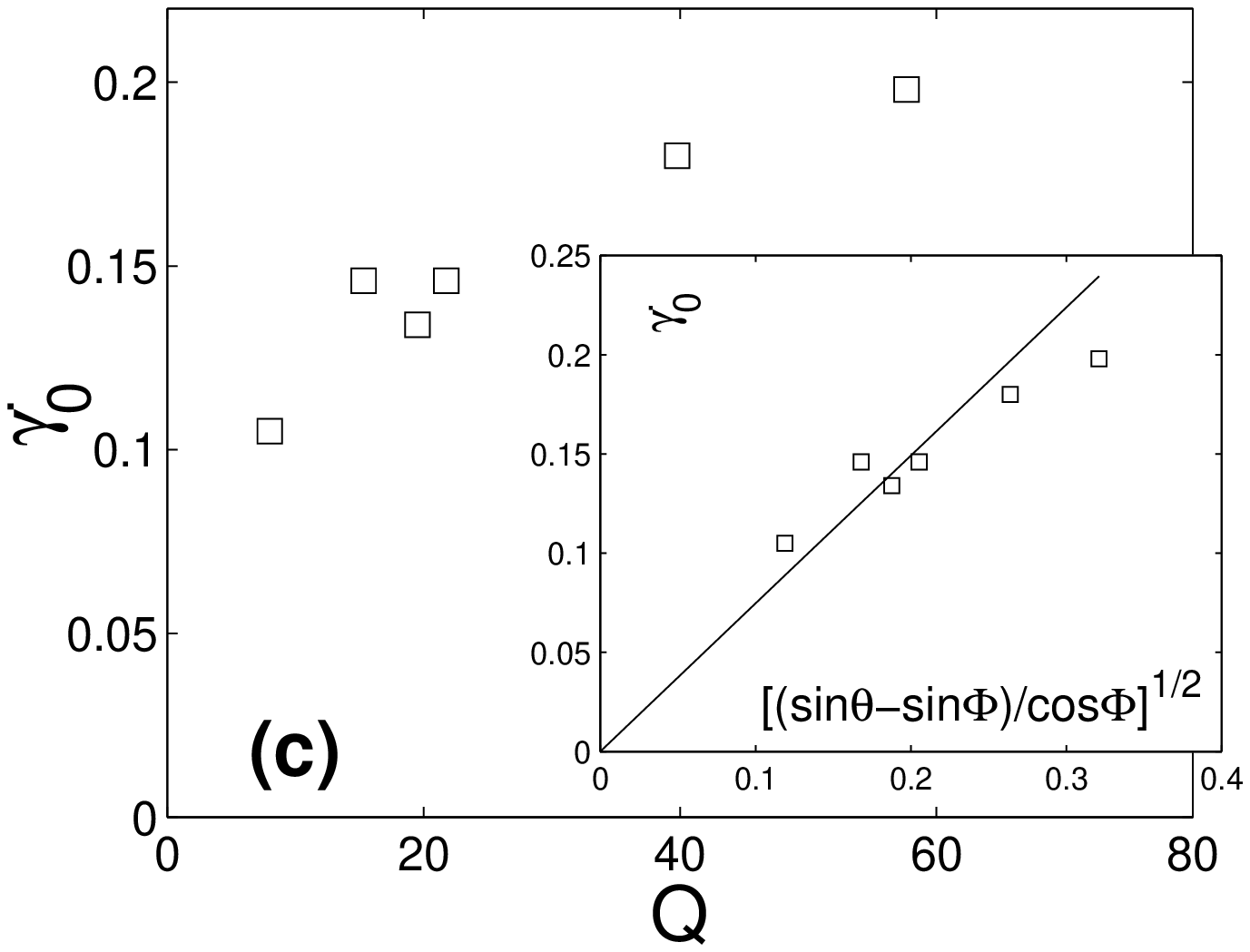}
\caption{(a) Velocity profiles at the center of the drum for various rotating speed $\Omega$ (b) Characteristic decay length $\lambda$ of the shear strain $\partial_z v_x$ as a function of the flow rate $Q$. Within the errorbars, $\lambda$ is constant, around $3$. (c) Constant shear rate $\dot{\gamma}_0$ within the flowing layer as a function of the flow rate $Q$. Inset: $\dot{\gamma}$ vs. $(\sin\theta-\sin\Phi)^{1/2}/\cos^{1/2}\Phi$ where the Coulomb friction angle $\Phi=17.4^\circ$ has been identified with the value $\mu_{eff}(Q=0)$ defined in Fig.~\ref{muvsq}b. The straight line is a linear fit $\dot{\gamma} = 0.75 (\sin\theta-\sin\Phi)^{1/2}/\cos^{1/2}\Phi$.}
\label{velocityvsq}
\end{figure}

Velocity profiles obtained in the center of the drum for various $Q$ are represented in Fig.~\ref{velocityvsq}a.
Apart from the flowing layer thickness $H$, the streamwise velocity profile at the center of the drum is characterized by two parameters, namely the characteristic decay length $\lambda$ of the exponential creep within the static phase and the constant shear rate $\dot{\gamma}_0$ within the flowing layer. Their evolution with respect to the flow rate $Q$ is reported in Fig~\ref{velocityvsq}b,c. Within the errorbars, $\lambda$ is independent of $Q$, of the order of $3\pm 0.3$. This behaviour is similar to what is reported in both 3D heap flows experiments~\cite{Komatsu01_prl} and 3D rotating drum experiments~\cite{Bonamy01_phd,Bonamy02_pof,CourrechdePont05_prl}, where $\lambda$ was found to be $\lambda \simeq 1.4$ and $\lambda \simeq 2.5$ respectively. 

In our numerical simulation, $\dot{\gamma}_0$ exhibits a weak dependence with $Q$ (see Fig.~\ref{velocityvsq}c). It ranges typically from $0.1$ to $0.25$ when $Q$ is made vary from $8$ to $58$. This dependency is compatible with the ones observed experimentally in 2D rotating drum by Rajchenbach~\cite{Rajchenbach00_ap}, who proposed that $\dot{\gamma}_0$ scales as $\dot{\gamma}_0 \propto (\sin\theta-\sin\Phi)^{1/2}/\cos^{1/2}\Phi$, where $\Phi$ refers to the Coulomb friction angle. Value of this angle can be estimated from the  variation of the mean flow angle with respect to $Q$ (see Fig. \ref{muvsq} and next section) and was found to be $\Phi=17.4^\circ$. Inset of Fig.~\ref{velocityvsq}c shows that the scaling proposed by Rajchenbach is compatible with our results. It is worth to mention that $\dot{\gamma}_0$ was found to be constant, around 0.5, independent of $Q$ in 3D experiments in Hele-Shaw drums~\cite{Bonamy02_pof,Gdrmidi04_epje}. This strongly suggests some non-trivial effect of either the lateral confinement or the flow dimension on the profile within the flowing layer. This will be explored in future 3D simulations of rotating drums.

\subsection{Flowing layer thickness and mean flow angle}

\begin{figure}[!tpb]
\centering
\includegraphics[height=0.6\columnwidth]{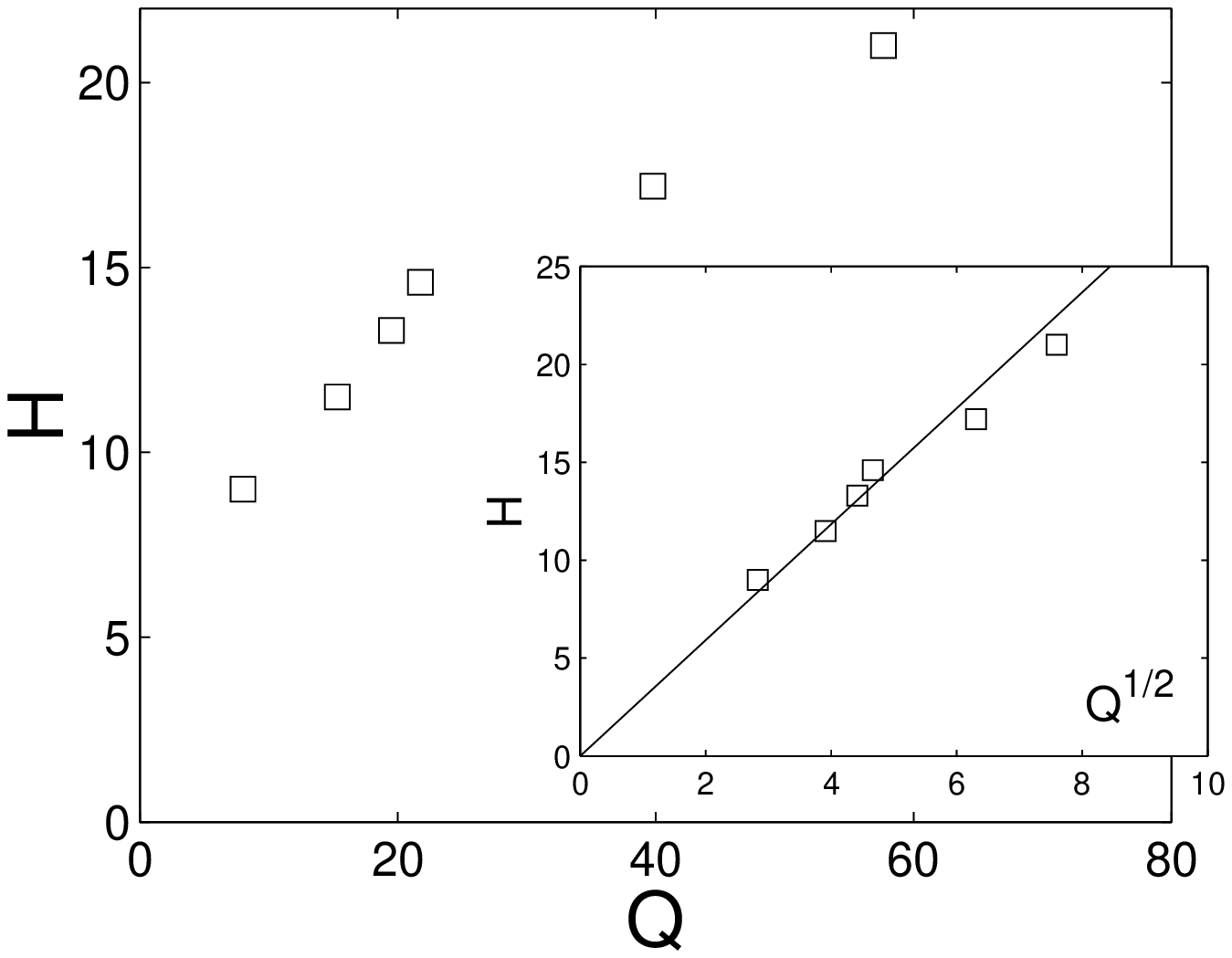}
\includegraphics[height=0.6\columnwidth]{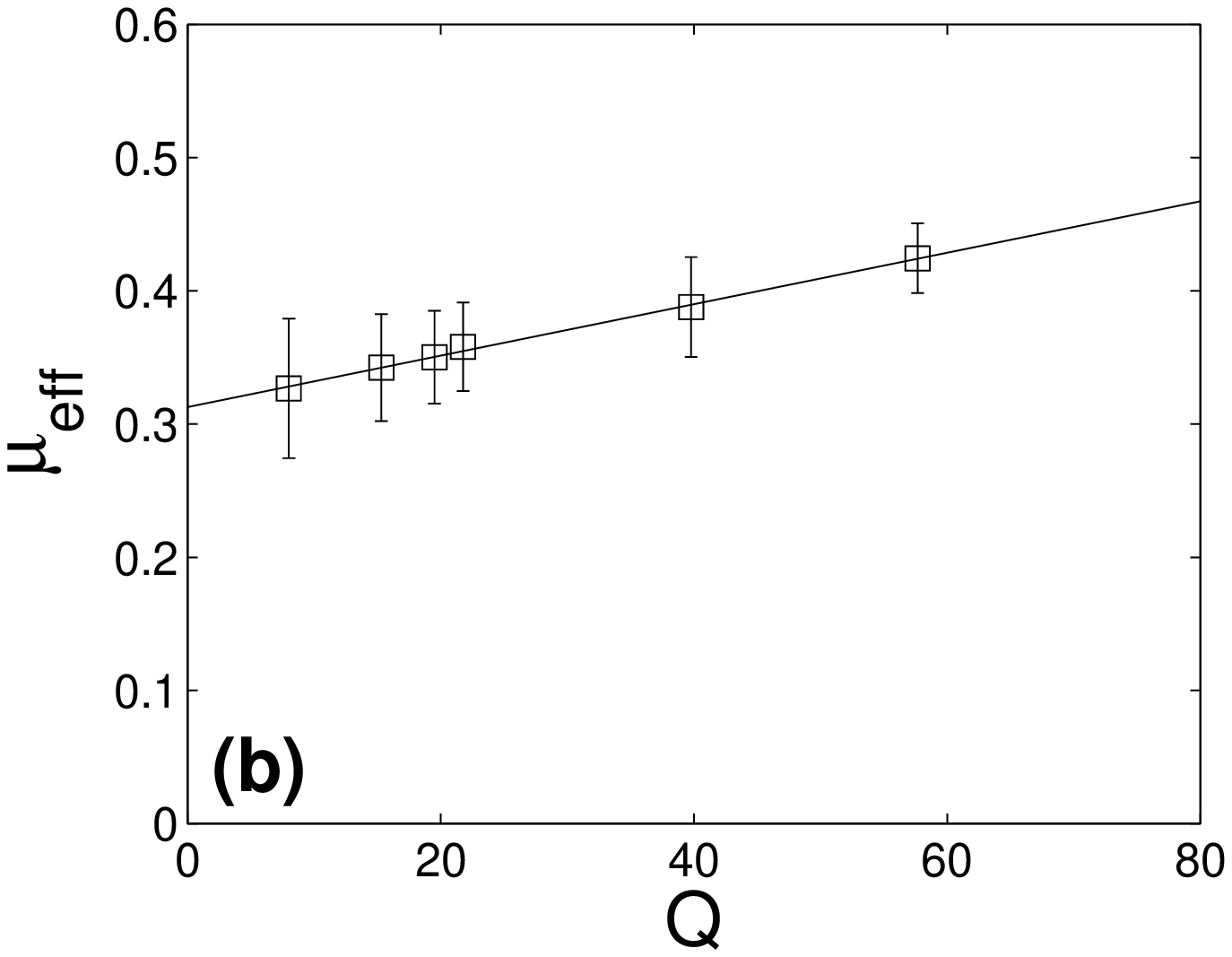}
\caption{(a) Flowing layer thickness $H$ at the center of the drum as a function of the flow rate $Q$. Inset, $H$ vs. $\sqrt{Q}$. The straight line is a linear fit $H = 3 \sqrt{Q}$. (b) Variation of the effective friction coefficient $\mu_{err}=\tan\theta$ of the surface flow with respect to $Q$ (non-dimensionalized units). The errorbars show the standard deviation over the sequence at constant $Q$. The straight line is a linear fit: $\mu_{eff}(Q)=0.31+1.9.10^{-3} Q$.}
\label{muvsq} 
\end{figure}

The thickness of the flowing layer $H$ is plotted as a function of the flow rate $Q$ in Fig. \ref{muvsq}a. As observed experimentally \cite{Bonamy02_pof,Gdrmidi04_epje}, $H$ scales as $\sqrt{Q}$, which is expected since the shear rate varies weakly within the flowing layer.

The mean flow angle $\theta$ can then be assimilated to an effective friction coefficient $\mu_{eff}=\tan\theta$ between the flowing layer and the static phase \cite{Douady99_epjb}. Its evolution with respect to the flow rate $Q$ is represented in Fig.~\ref{muvsq}. The effective friction coefficient is found to increase with $Q$. Similar increasing was observed experimentally, -~at a much larger scale~-. It was attributed to wall effects~\cite{Bonamy02_pof,Gdrmidi04_epje} since this dependency was observed to be weaker when the drum thickness is increased~\cite{Bonamy01_phd,Gdrmidi04_epje}. No such wall effects can be invoked in the present study. In other words, part of the increase of the effective friction with flow rate cannot be induced by wall friction contrary to what was suggested in~\cite{Bonamy02_pof,Gdrmidi04_epje} and should be found in the granular flow rheology.

\subsection{Angular velocity profiles}

\begin{figure}[!tpb]
\centering
\includegraphics*[height=0.6\columnwidth]{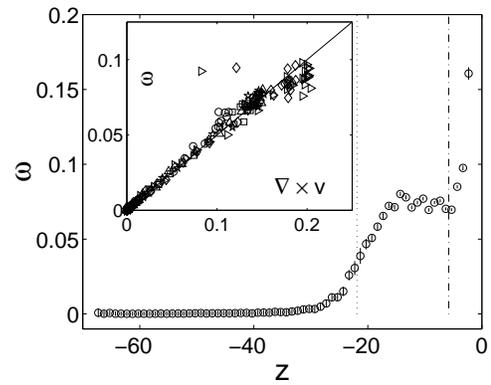}
\caption{Mean angular velocity profile $\omega(z)$ (averaged over 400 frames) at the center of the drum obtained for $\Omega=6\un{rpm}$. The errorbars correspond to a $99\%$ confident interval. The vertical mixed line (resp. the vertical dotted line) show the free surface boundary as defined in Fig.~\ref{comp6rpm} (resp. the static/flowing interface as defined in Fig.~\ref{vit6rpm}). Inset, $\omega$ vs vorticity $\nabla \times {\bf v}=\partial_z v_x$ for $\Omega=2\un{rpm}$, ($o$), $\Omega=4\un{rpm}$ ($\star$), $\Omega=5\un{rpm}$ ($\square$), $\Omega=6\un{rpm}$ ($\triangle$), $\Omega=10\un{rpm}$ ($\diamond$), $\Omega=15\un{rpm}$ ($\triangleright$). The straight line is given by $\omega = \frac{1}{2}\nabla \times {\bf v}$.}
\label{spin6rpm}
\end{figure}

A typical mean angular velocity profile $\omega(z)$ has been represented in Fig.~\ref{spin6rpm}. It is interesting to plot $\omega$ with respect to the vorticity $\nabla \times \mathbf{v}=\partial_z v_x$ (see Inset of Fig.~\ref{spin6rpm}). There is a clear relationship between these two quantities: $\omega=\frac{1}{2}\nabla \times \mathbf{v}$ in {\em the whole packing} independently of $Q$. This relationship is analogue to the one obtained in classical hydrodynamics where the mean rotating speed of the particles is equal to half the vorticity. Such relationship was observed in Molecular Dynamics simulations of dilute granular flows~\cite{Campbell85_jam,Lun91_jfm}, but expected to fail at higher volume fraction~\cite{Campbell86_aa,Campbell86_jfm}. In this latter case, grains were expected to organize in layers the grains of which rotate in the same direction. This would decrease the mean angular velocity of the grains, and $\omega$ would be smaller than $\frac{1}{2} \nabla \times \mathbf{v}$. In our numerical simulation, such behaviour is not observed, which suggests that the grains spins do not organize in layer despite the high density of the flow.

\section{Stress analysis}

\subsection{Stress tensor profile}

Let us now look at stress profiles - that cannot be measured experimentally. The stress tensor $\mathbf{\sigma}$ is the sum of three contributions: $\mathbf{\sigma}=\mathbf{\sigma}^c+\mathbf {\sigma}^k+\mathbf{\sigma}^r$ where $\mathbf{\sigma}^c$, $\mathbf{\sigma}^k$ and $\mathbf{\sigma}^r$ refer to the contact, kinetic and rotational components of the stress tensor respectively. In our dense free surface flows, $\mathbf{\sigma}^k$ and $\mathbf{\sigma}^r$ are found to be negligible with regard to $\mathbf{\sigma}^c$. One can thus assume that $\mathbf{\sigma} \simeq \mathbf{\sigma}^c$.  

\begin{figure*}[!tpb]
\centering
\includegraphics*[height=0.45\columnwidth]{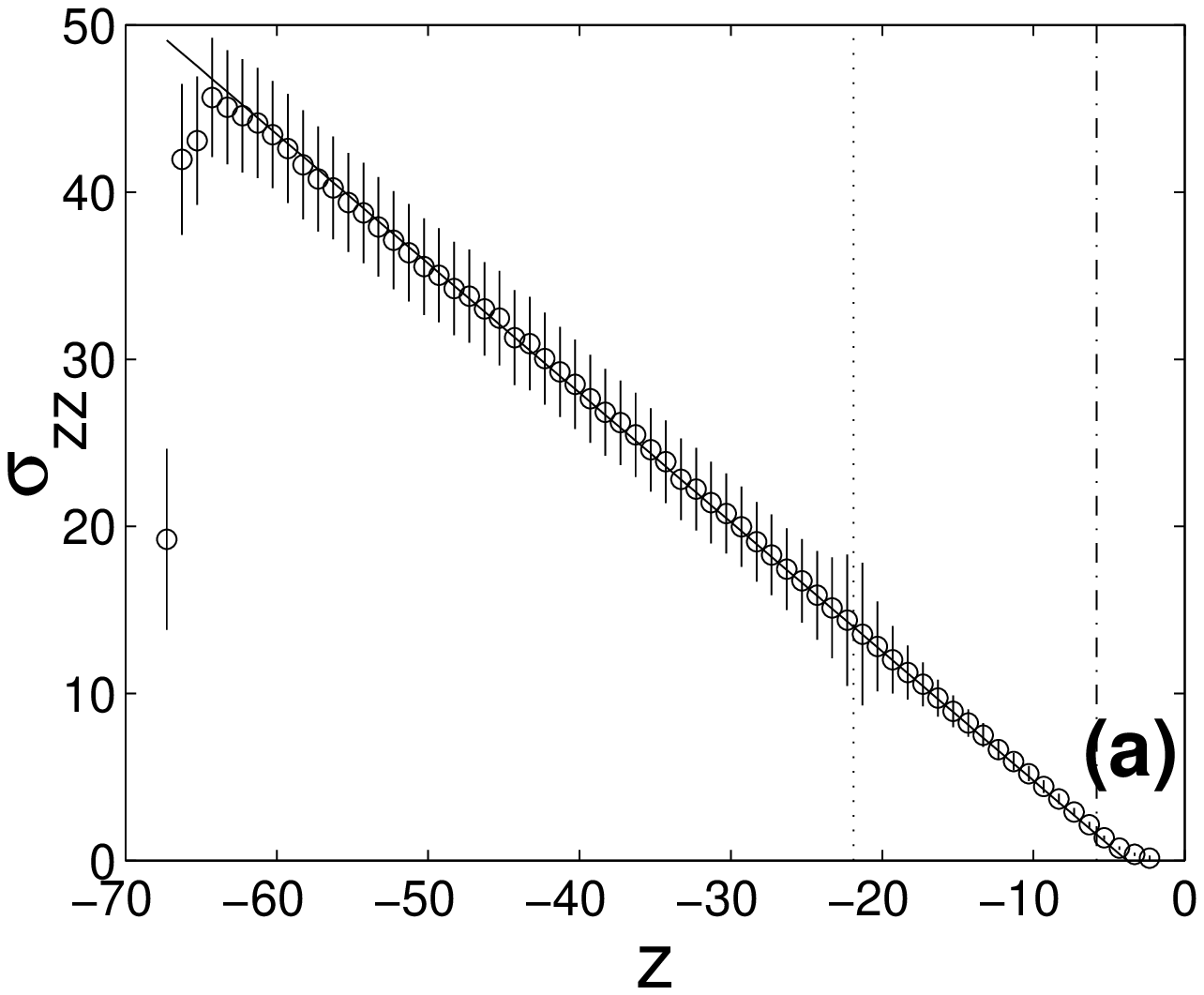}
\includegraphics*[height=0.45\columnwidth]{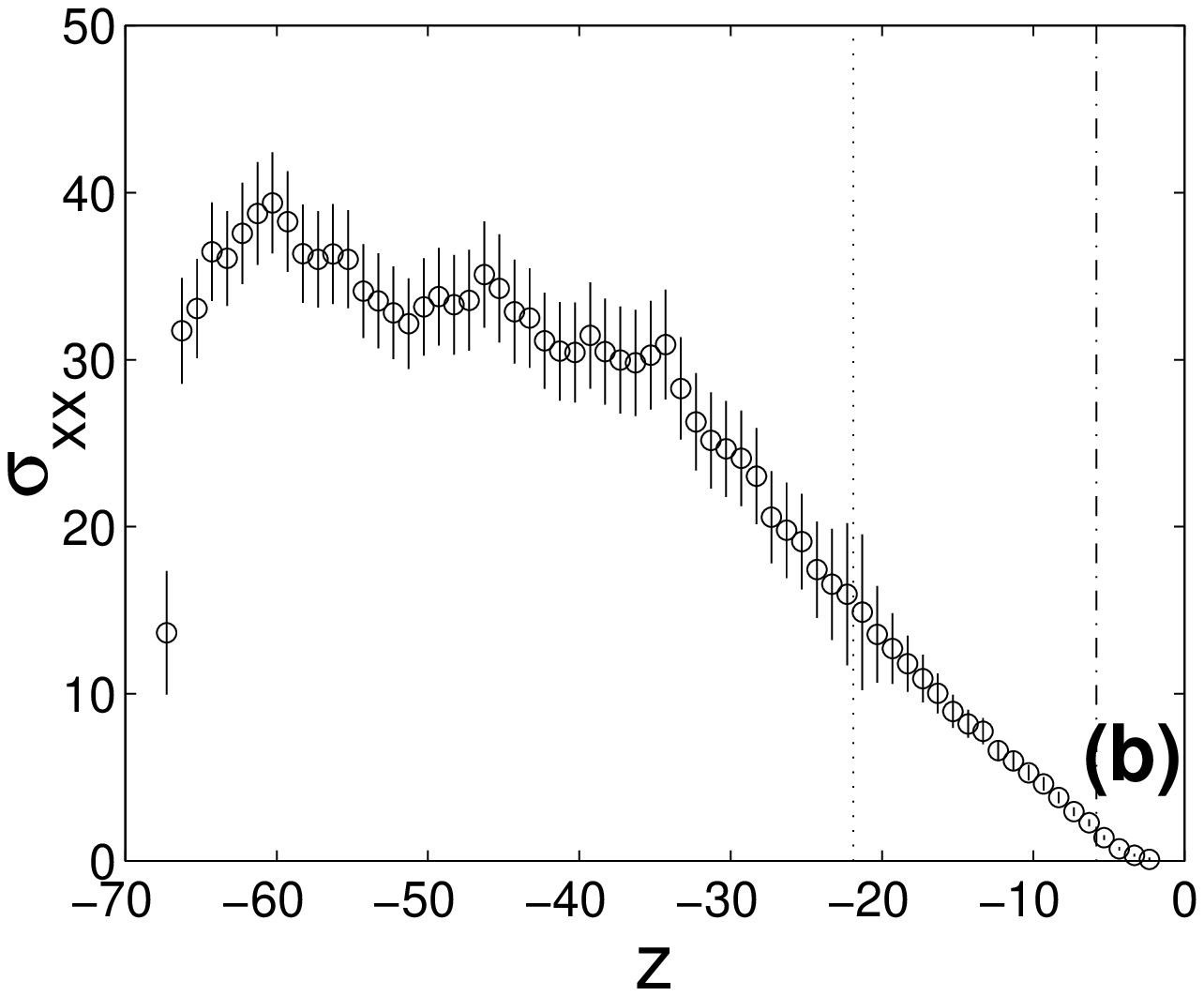}
\includegraphics*[height=0.45\columnwidth]{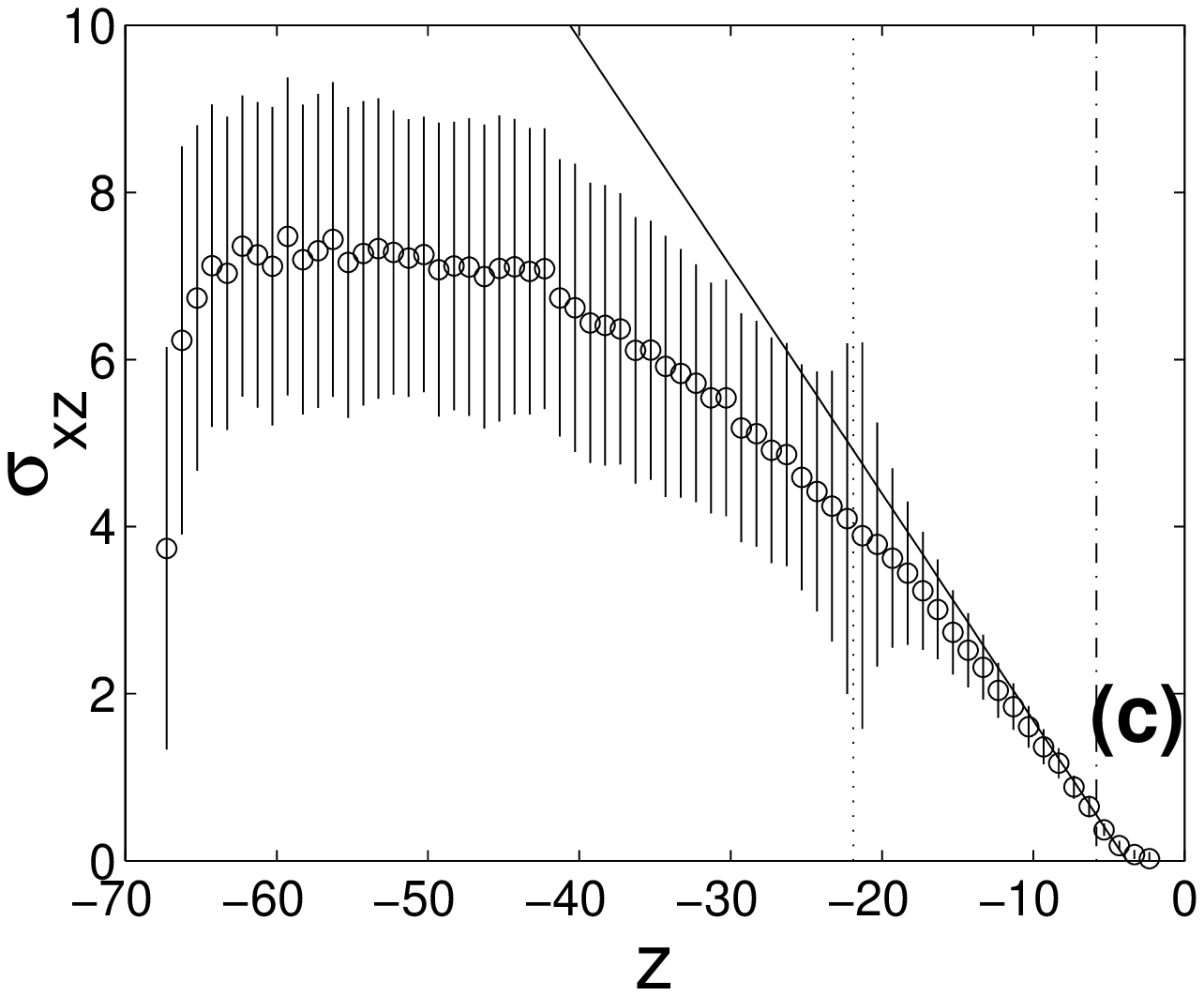}
\caption{Profiles of the three independent components of the static stress tensor $\sigma_{zz}$ (a), $\sigma_{xx}$ (b) and $\sigma_{xz}$ (c) for $\Omega=6\un{rpm}$. The errorbar correspond to a $99\%$ confident interval. In subfigure (a), (resp. subfigure (c)) the straight line is given by $\sigma_{zz}=-\nu^{RCP}\cos\theta$ (resp. by $\sigma_{xz}=-\nu^{RCP}\sin\theta$) as expected from Cauchy equations for an homogenous surface incompressible flow of volume fraction $\nu^{RCP} = 0.82$. For $\Omega=6\un{rpm}$, $\theta$ was measured to be $\theta=19.7^\circ$. The vertical mixed lines and the vertical dotted lines show the free surface boundary and the static/flowing interface as defined in Fig.~\ref{comp6rpm} and Fig.~\ref{vit6rpm} respectively.}
\label{Sig6rpm} 
\end{figure*}

Components of the contact stress tensor associated to each bead has been computed for each snapshot of each numerical experiment (see Sec.\ref{setup}). The profile of the continuum value of each component of the contact stress tensor - and consequently the total stress tensor - $\sigma_{\alpha\beta}(z)$ is then defined over the elementary slice $\Sigma$ located in the center of the drum (see Fig. \ref{snapshot}) according to the same procedure used to calculate velocity and volume fraction profile. The tensor $\mathbf{\sigma}$ is found to be symmetric, \ie $\sigma_{xz} = \sigma_{zx}$. For 2D surface flows, it is thus defined by three independent components $\sigma_{xx}$, $\sigma_{xz}$ and $\sigma{zz}$. Typical profile of these components with respect to the depth $z$ at the center of the drum are represented in Fig. \ref{Sig6rpm}.

Shapes of these profiles are quite surprising. In the rotating frame $\Re$, velocity and volume fraction profiles were found to be invariant along infinitesimal translation along ${\bf e}_x$ within the elementary slice $\Sigma$. In other words, for $x=0$, one gets $\partial\nu/\partial x \simeq \partial {\bf v}/\partial x\simeq 0$. More generally, it is commonly assumed that at the center of the drum, the $x$ derivative of the stress tensor vanishes \cite{Gdrmidi04_epje,Rajchenbach03_prl,Rajchenbach00_ap,Bonamy01_phd,Bonamy02_prl}. The Cauchy equations would then read:

\begin{equation}
\begin{array}{ll}
(a) \quad & \frac{\partial \sigma_{xz}}{\partial z} = -\nu\sin\theta \\
(b) \quad & \frac{\partial \sigma_{zz}}{\partial z} = \nu\cos\theta +\nu\Omega v_x +\nu\Omega^2 z
\end{array}
\label{cauchytot}
\end{equation}

\noindent where $\theta$ denotes the mean flow angle. The second right-handed term of Eq.~\ref{cauchytot}b is the Coriolis term. This term is maximum at the free surface where it reaches $15\%$ of the first right-handed gravity term \eg for $\Omega=6\un{rpm}$. the last right-handed term of Eq.~\ref{cauchytot}b is the centrifugal term. This term is maximum at the drum boundary where it reaches $1\%$ of the first right-handed gravity term \eg for $\Omega=6\un{rpm}$. Finally, inertial effects can be neglected and the Cauchy equations for pure steady homogenous flows would come down to:

\begin{equation}
\begin{array}{ll}
(a) \quad & \frac{\partial \sigma_{xz}}{\partial z} = -\nu\sin\theta \\
(b) \quad & \frac{\partial \sigma_{zz}}{\partial z} = -\nu\cos\theta
\end{array}
\label{cauchysimple}
\end{equation}

\noindent and, since the volume fraction $\nu$ is almost constant, close to the random close packing value $\nu^{RCP}=0.82$:

\begin{equation}
\begin{array}{ll}
(a) \sigma_{xz}(z) = -z \nu^{RCP}\sin\theta \\
(b) \sigma_{zz}(z) = -z \nu^{RCP}\cos\theta
\end{array}
\label{stresssimple}
\end{equation}

\begin{figure*}[!tpb]
\centering
\includegraphics*[height=0.45\columnwidth]{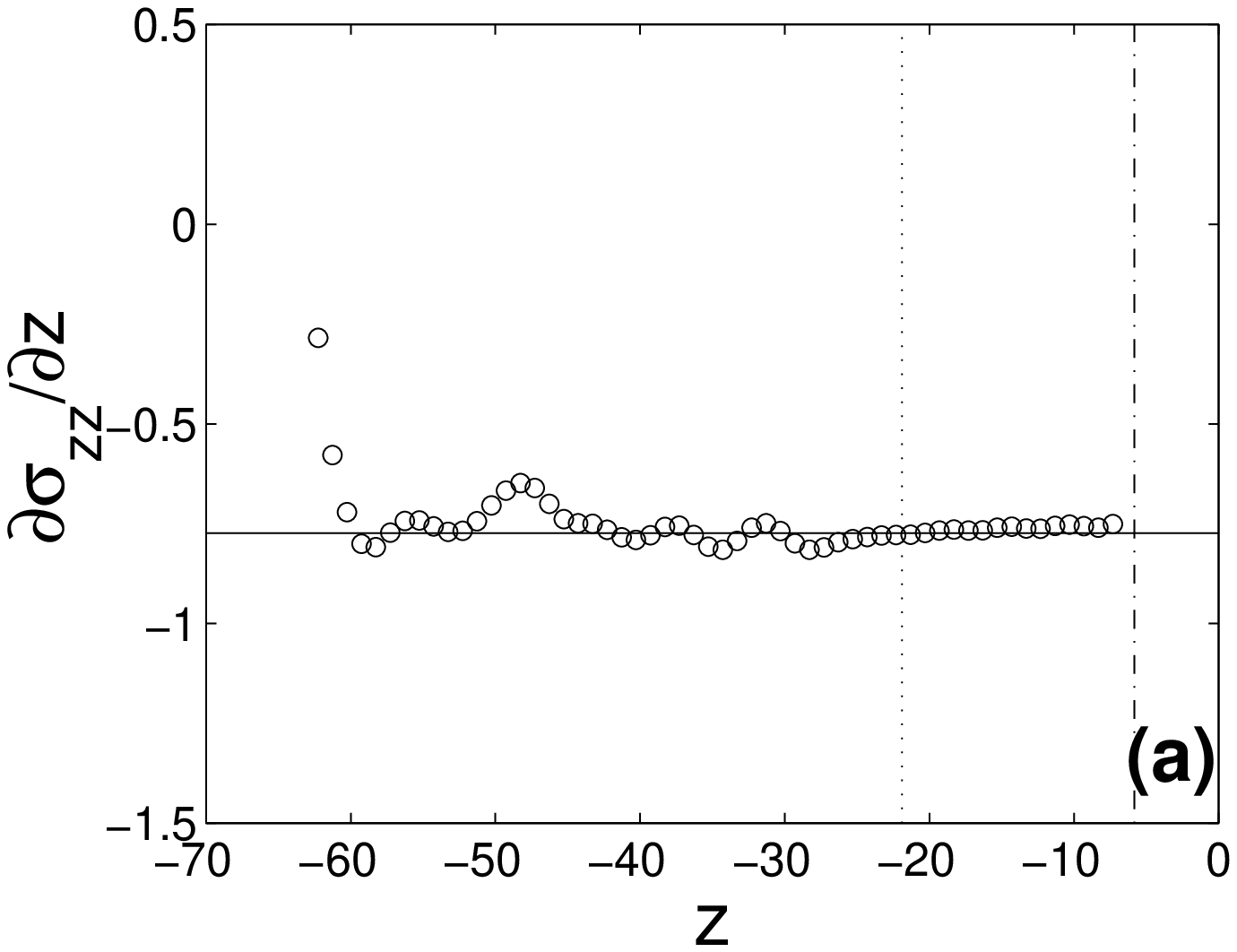}
\includegraphics*[height=0.45\columnwidth]{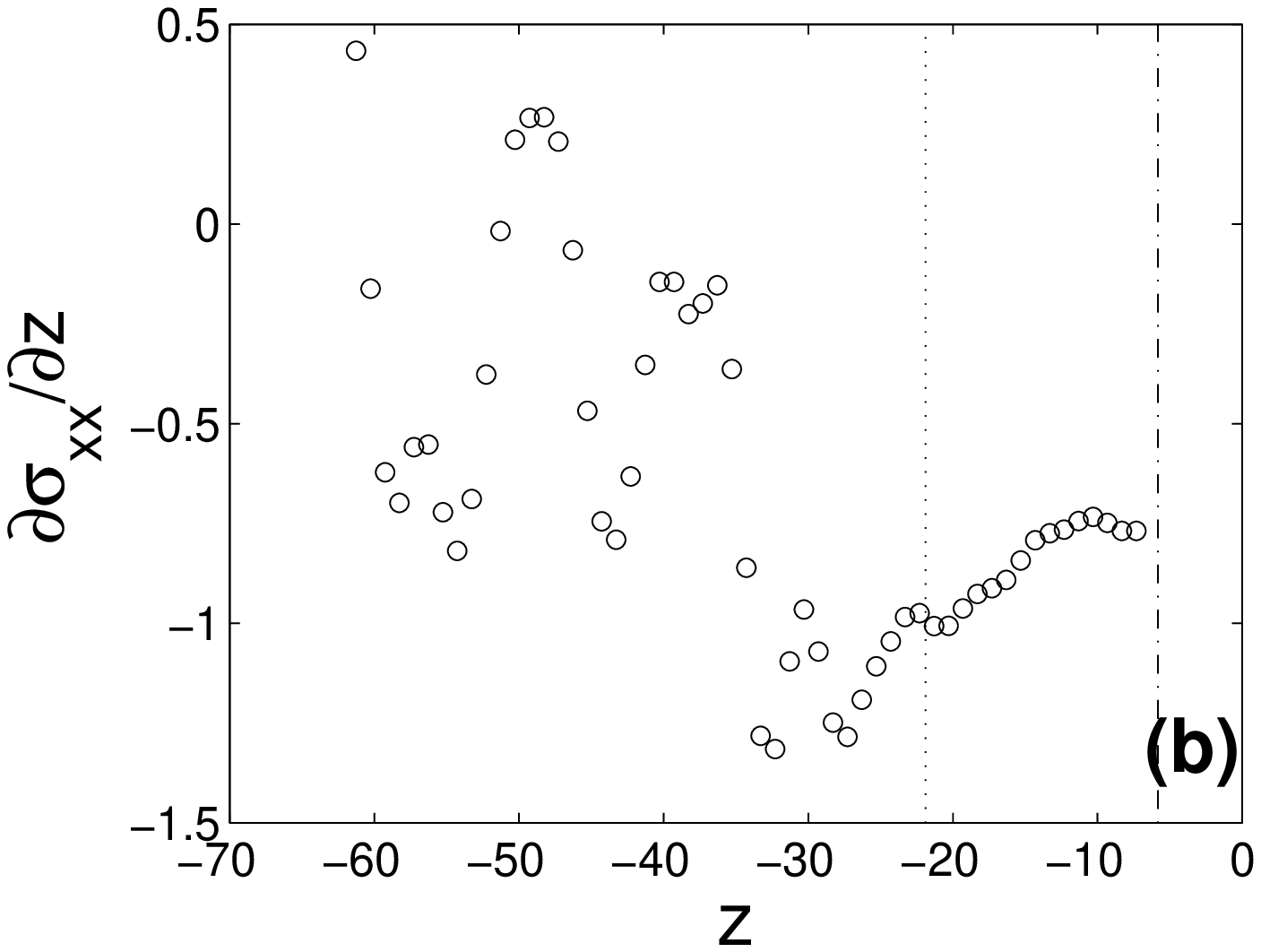}
\includegraphics*[height=0.45\columnwidth]{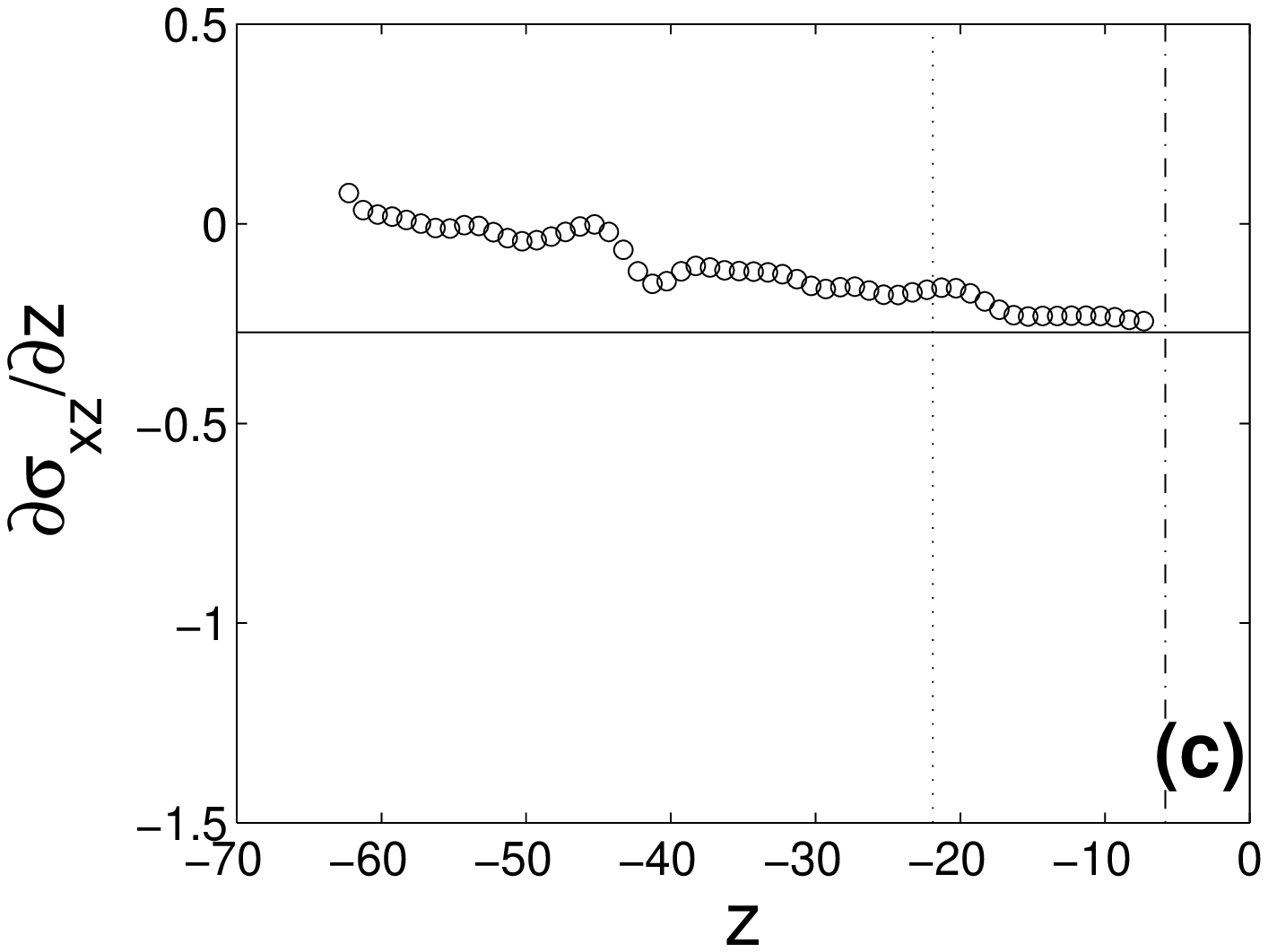}
\includegraphics*[height=0.45\columnwidth]{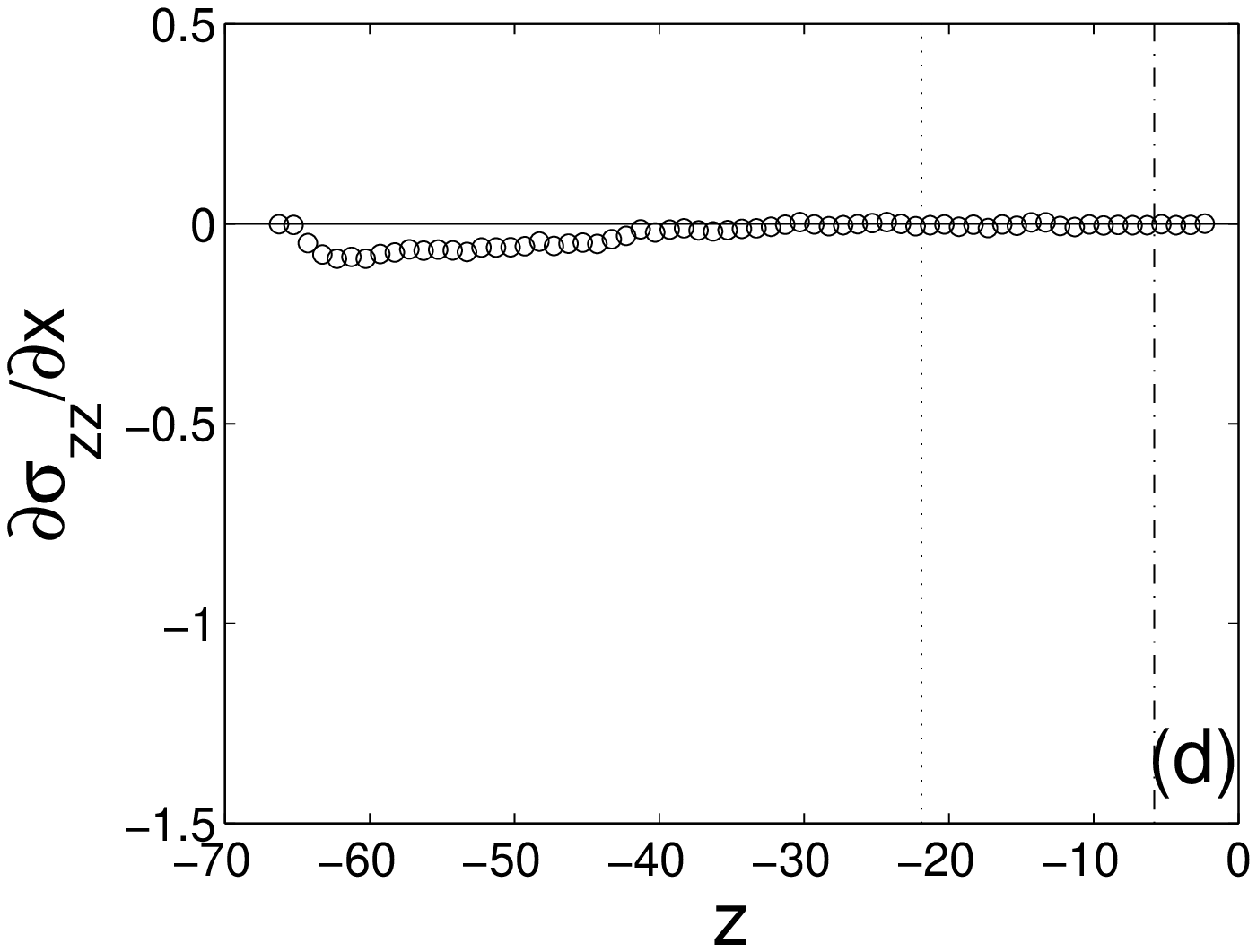}
\includegraphics*[height=0.45\columnwidth]{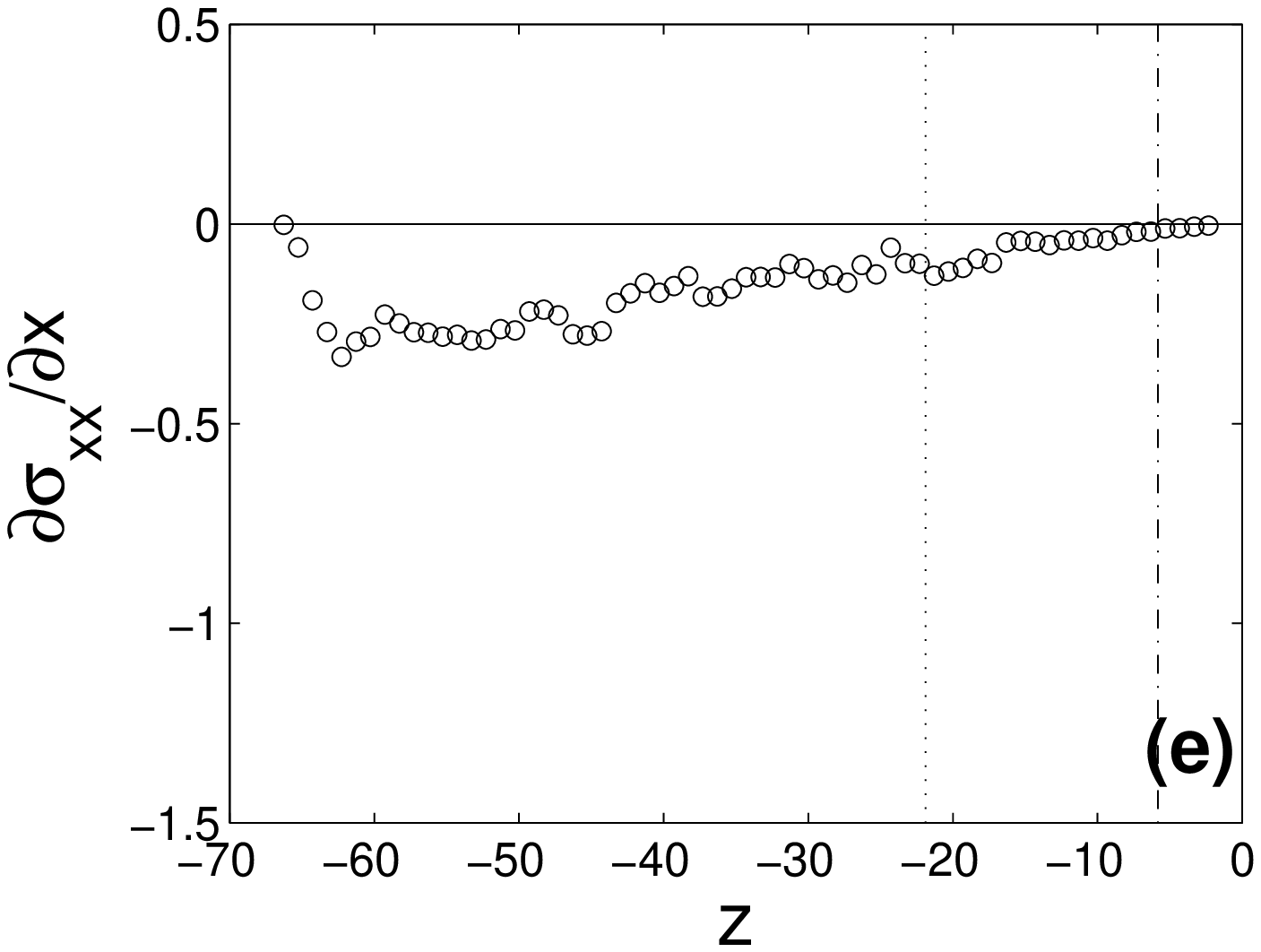}
\includegraphics*[height=0.45\columnwidth]{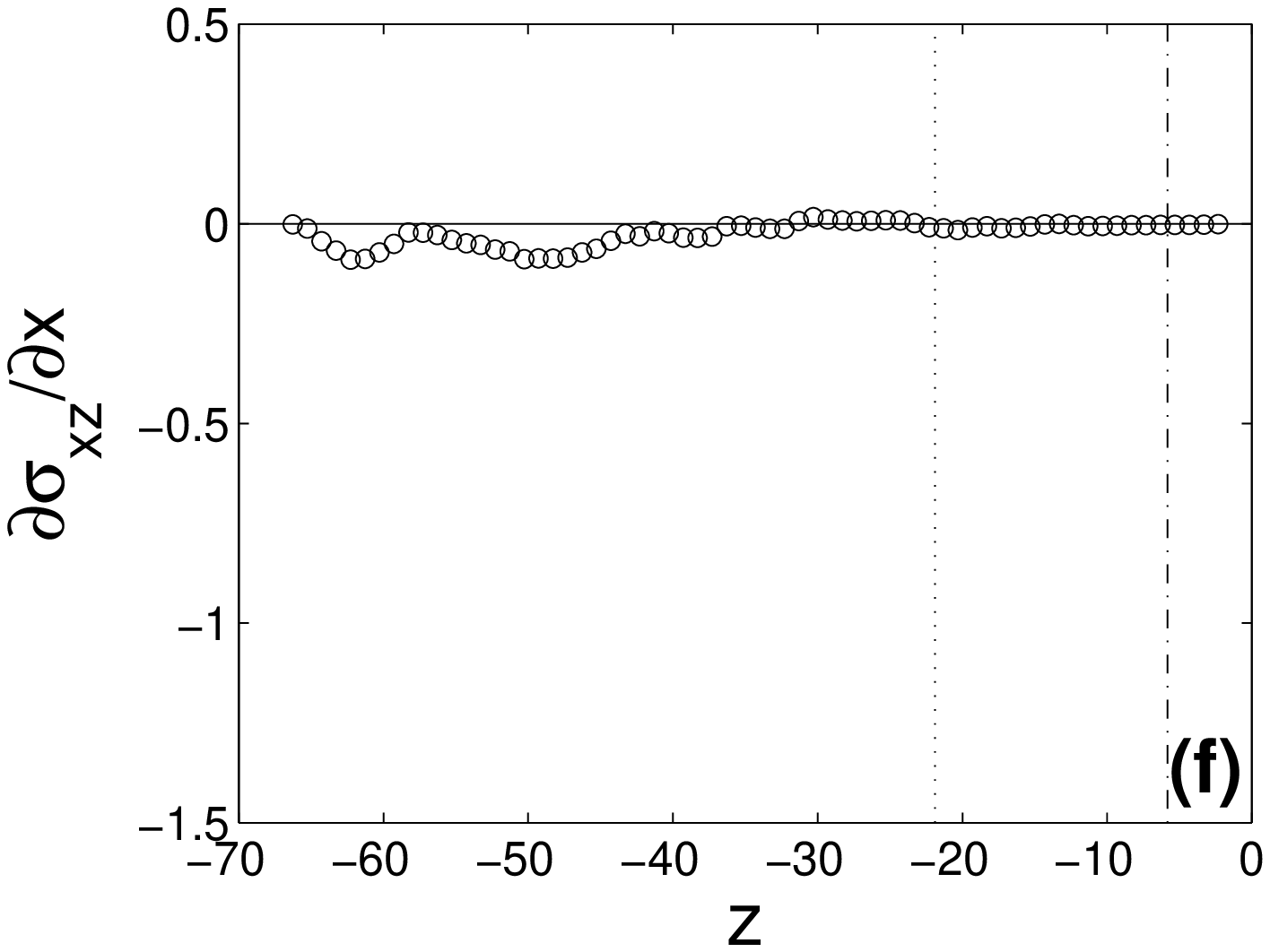}
\caption{Top: Profiles of the 6 independent component of the contact stress tensor gradient, namely $\partial \sigma_{zz}/\partial z$ (a), $\partial \sigma_{xx} / \partial z$ (b), $\partial \sigma_{xz}/\partial z$ (c), $\partial \sigma_{zz}/\partial x$ (d), $\partial \sigma_{xx} / \partial x$ (e) and $\partial \sigma_{xz}/\partial x$ (f) for $\Omega=6\un{rpm}$. The errorbars correspond to a $99\%$ confident interval. In subfigure (a), (c), (d) and (f), the horizontal straight line is given by $\partial\sigma_{zz}/\partial z=-\nu^{RCP}\cos\theta$, $\partial\sigma_{xz}/\partial z=-\nu^{RCP}\sin\theta$, $\partial\sigma_{zz}/\partial x=0$ and $\partial\sigma_{xz}/\partial x=0$ respectively, as expected from Cauchy equations for an homogenous surface incompressible flow of volume fraction $\nu^{RCP} = 0.82$ and mean flow angle $\theta=19.7^\circ$ as measured for $\Omega=6\un{rpm}$. The vertical mixed lines and the vertical dotted lines show the free surface boundary and the static/flowing interface as defined in Fig.~\ref{comp6rpm} and Fig.~\ref{vit6rpm} respectively.}
\label{dSdR} 
\end{figure*}

\noindent These predictions were compared to the measured profiles (Fig. \ref{Sig6rpm}). The measured profile $\sigma_{zz}$ fits well with Eq.~\ref{stresssimple}b. However, $\sigma_{xz}$ departs from Eq.~\ref{stresssimple}a within the static phase. To understand this discrepancy, one looks at the gradient of the stress tensor (see Fig.~\ref{dSdR}). The $x$-derivative of the various components were calculated by translating the elementary slice $\Sigma$ of an increment $\delta x=5$ from one side to the other of the reference position $x=0$. We checked that the obtained values do not depend on $\delta x$. Both $\partial \sigma_{zz}/\partial x$ and $\partial \sigma_{xz} / \partial x$ vanish within $\Sigma$ at the center of the drum. However $\partial \sigma_{xx}/\partial x$ does not. In other words, steady surface flows in rotating drums cannot be considered as quasi-homogenous even at the center of the drum. The Cauchy equations should then read:

\begin{equation}
\begin{array}{ll}
(a) \quad & \frac{\partial \sigma_{xx}}{\partial x} + \frac{\partial \sigma_{xz}}{\partial z} = -\nu\sin\theta \\
(b) \quad & \frac{\partial \sigma_{zz}}{\partial z} =- \nu\cos\theta
\end{array}
\label{cauchydrumsimple}
\end{equation}

\noindent This may explain the slight discrepancies observed between homogenous steady heap surface flows and steady surface flows in rotating drum (see \eg \cite{Bonamy03_epl} for related discussions).

\section{Constitutive laws}

\subsection{Inertial number $I$}

It was recently suggested \cite{Dacruz04_proc,Dacruz04_phd,Gdrmidi04_epje,Jop05_jfm} that the shear state of a dense granular flow can be characterized through a dimensionless number $I$, referred to as the inertial number, defined as:

\begin{equation} 
I=\frac{\partial_z v_x}{\sqrt{\sigma_{zz}}}
\label{I}
\end{equation}

\noindent This parameter can be regarded as the ratio between the typical time of deformation $1/\partial_z v_x$ and the typical time of confinement $1/\sqrt{\sigma_{zz}}$ \cite{Gdrmidi04_epje}.

\begin{figure}[!tpb]
\centering
\includegraphics*[height=0.6\columnwidth]{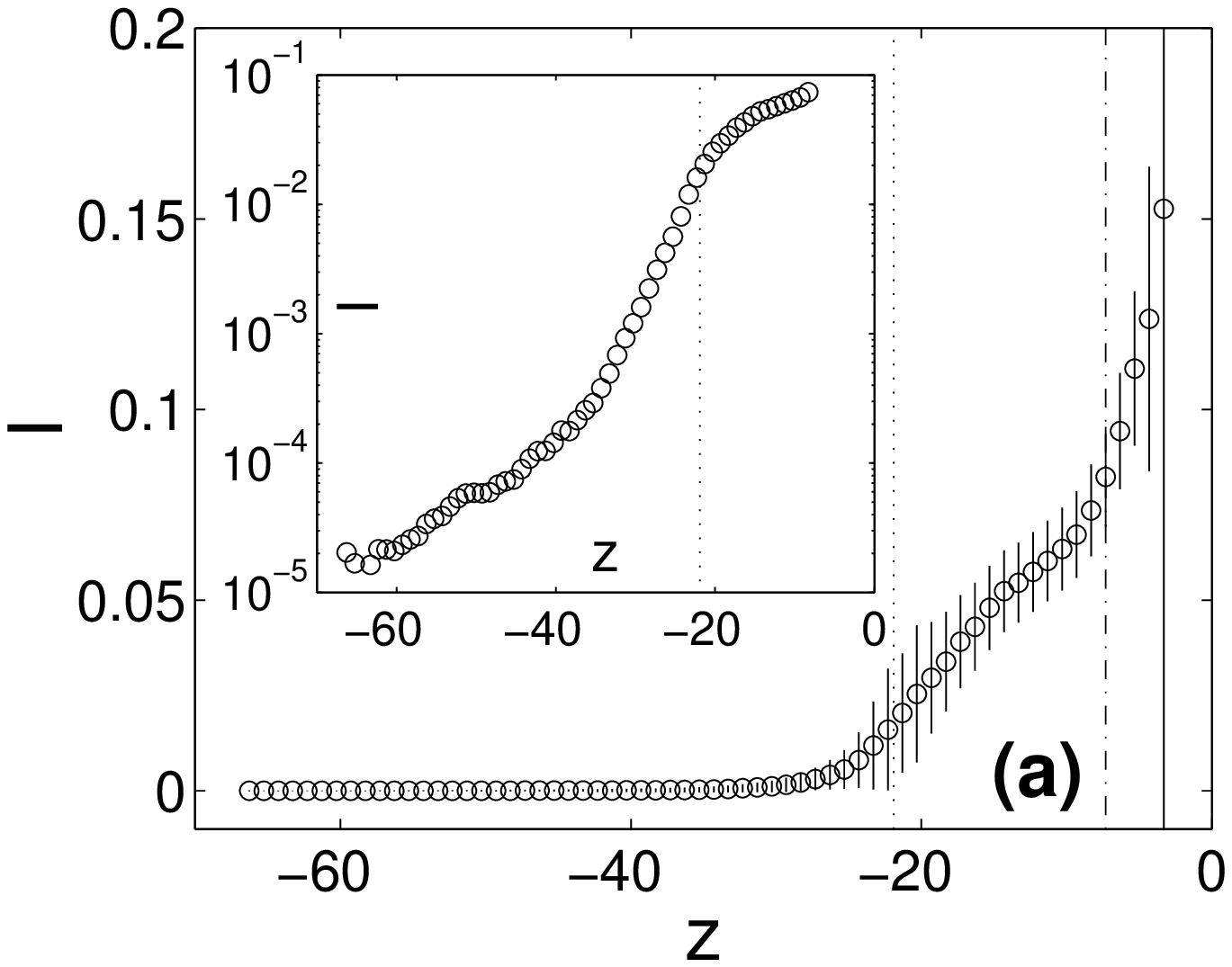}
\includegraphics*[height=0.6\columnwidth]{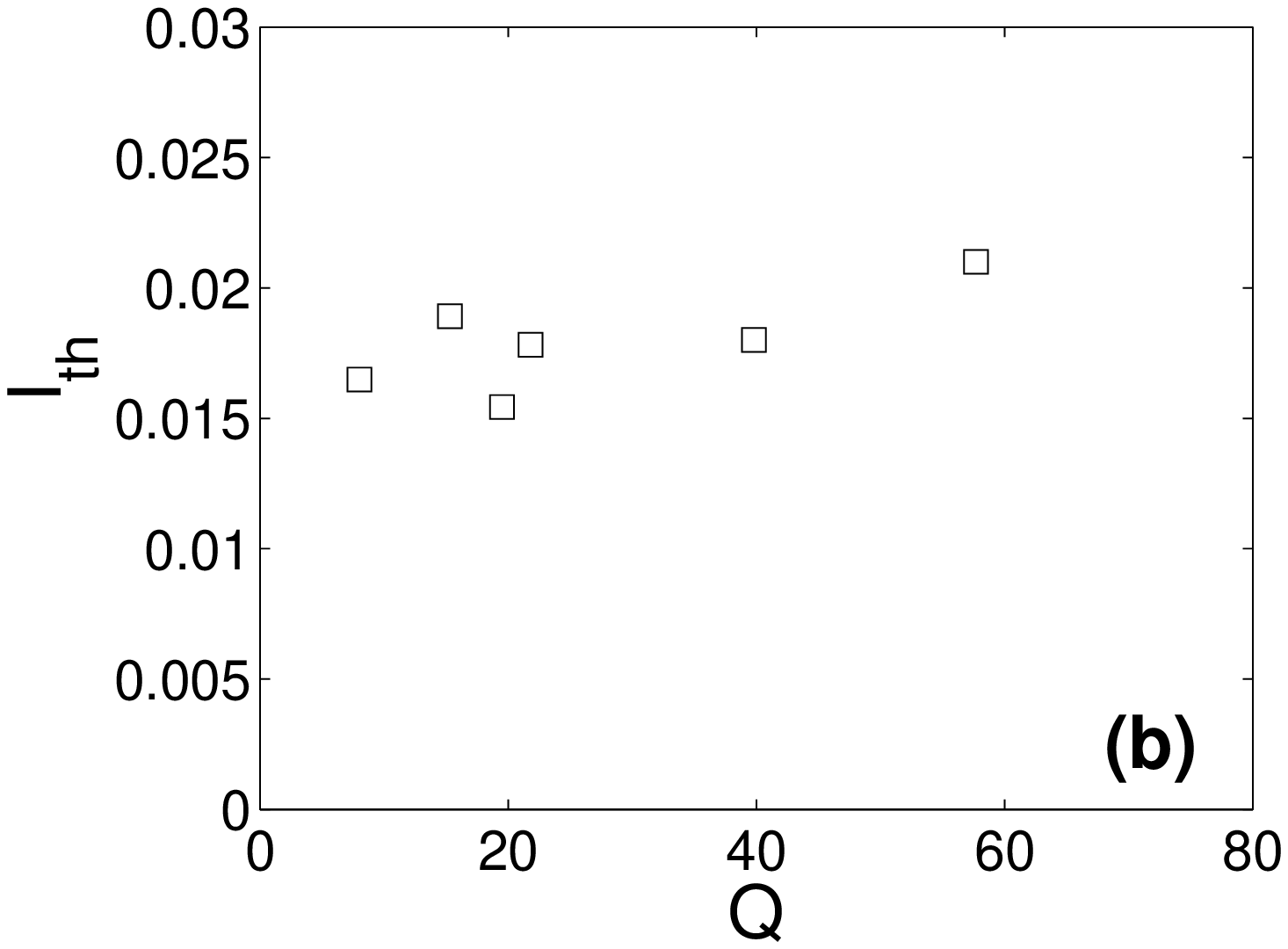}
\caption{(a): Profiles of the inertial parameter $I(z)$ for $\Omega=6\un{rpm}$. The errorbar correspond to a $99\%$ confident interval. The vertical mixed lines and the vertical dotted lines show the free surface boundary and the static/flowing interface as defined in Fig.~\ref{comp6rpm} and Fig.~\ref{vit6rpm} respectively. (b): Variation of the value $I_{th}$ of the inertial number at the static/flowing interface with respect to the flow rate $Q$.}
\label{I6rpm} 
\end{figure}

A typical profile of the inertial number $I$ is plotted in Fig. \ref{I6rpm}a. This non-dimmensionalized parameter was shown to be the relevant parameter to account for the transition from the quasi-static regime to the dense inertial regime in plane shear configuration, annular shear and inclined plane configuration \cite{Dacruz04_proc,Dacruz04_phd,Gdrmidi04_epje}. Therefore, it is natural to consider $I$ as the relevant parameter to describe the transition from the quasi-static phase and the flowing layer in the surface flow geometry. To check this assumption, we determine the value $I_{th}$ of the inertial number at the interface between the static phase/flowing layer interface - defined by extrapolating the linear velocity profile of the flowing phase to zero (see Fig.~\ref{vit6rpm})- for all our numerical experiments carried out at various $\Omega$. Variations of $I_{th}$ as a function of $Q$ is represented in Fig. \ref{I6rpm}b. This threshold is found to be constant, equal to:

\begin{equation} 
I_{th} \simeq 1.8 . 10^{-2}
\label{Ith}
\end{equation}

\noindent which provides a rather strong argument to consider this non-dimensionalized parameter as the relevant one to describe surface flows.

\subsection{Rheology}

Now that a relevant parameter describing the local shear state of the flow has been proposed, one can discuss in more detail the flow rheology. As a first guess, it is tempting to consider {\em local} constitutive laws relating the components of the stress tensor to $I$ through a one-to-one relation. In this case, dimensional analysis leads to:

\begin{equation} 
\sigma_{xz}/\sigma_{zz} = \mu(I),\quad \sigma_{xx}/\sigma_{zz} = k(I)
\label{eqrheo}
\end{equation}

\begin{figure}[!tpb]
\centering
\includegraphics*[height=0.6\columnwidth]{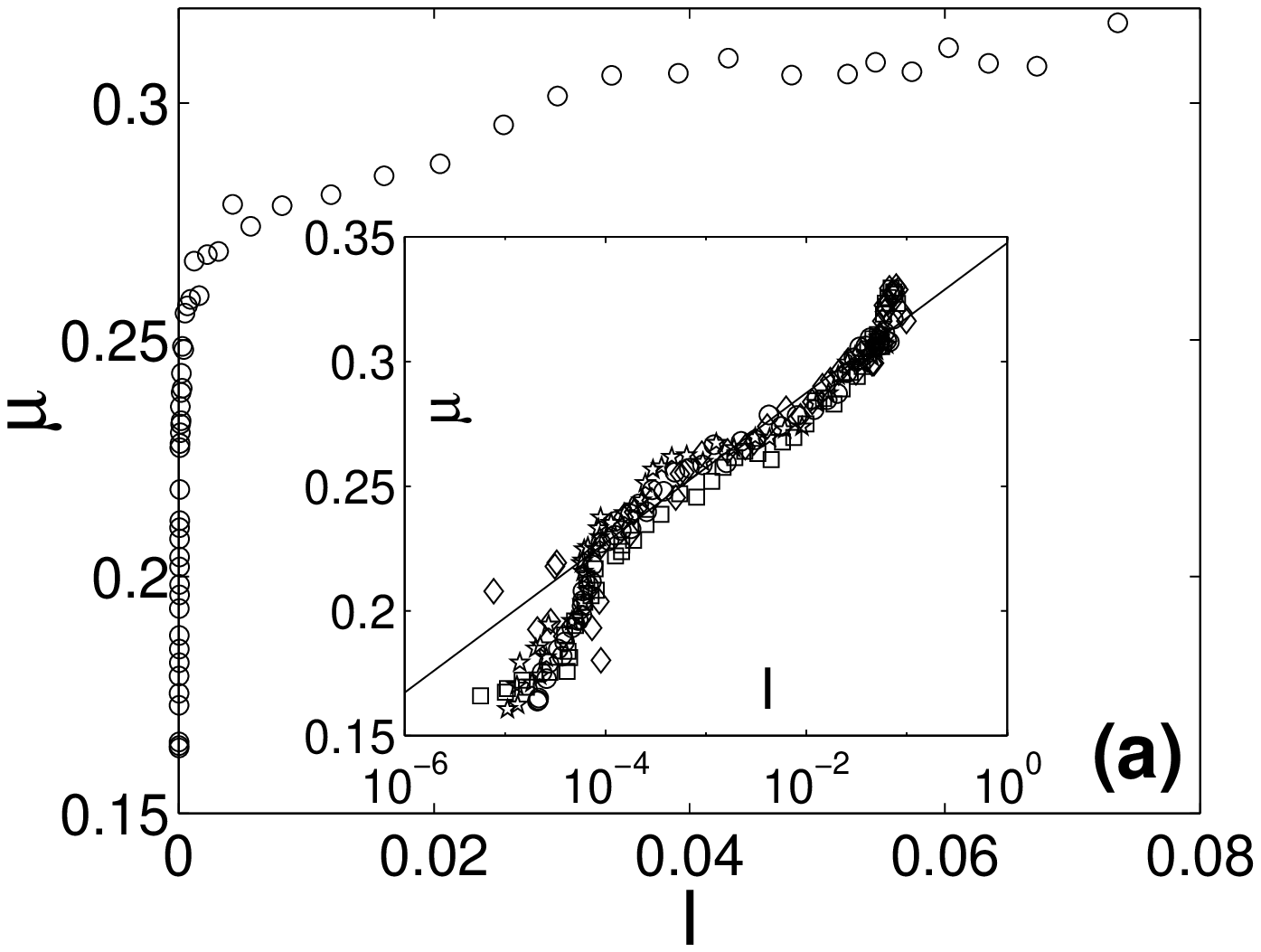}
\includegraphics*[height=0.6\columnwidth]{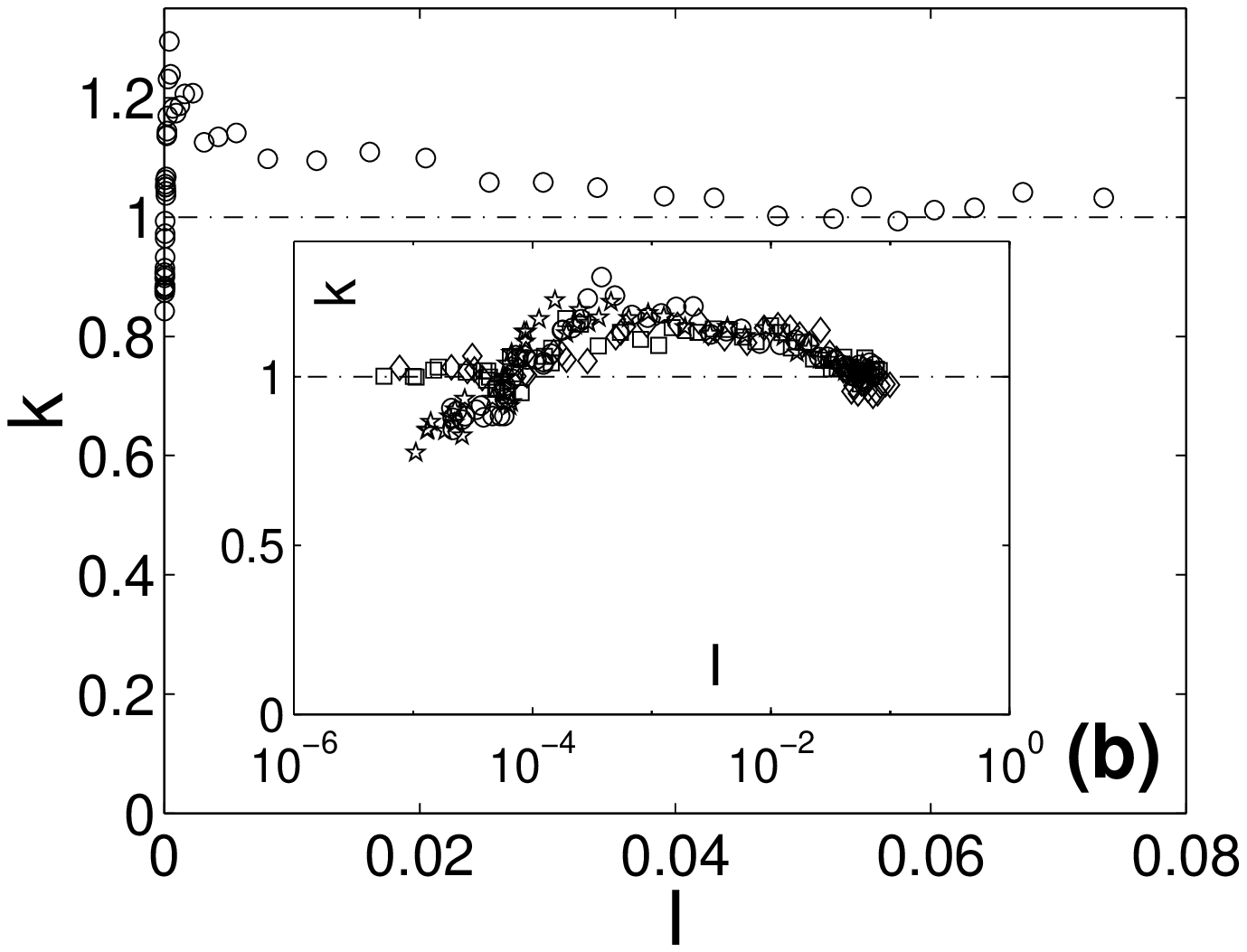}
\caption{(a): Variation of the effective friction coefficient $\mu=\sigma_{xz}/\sigma_{zz}$ as a function of the inertial number $I$ for $\Omega=6\un{rpm}$. Inset: Variation of $\mu$ as a function of $I$ for $\Omega=2\un{rpm}$, ($o$), $\Omega=4\un{rpm}$ ($\star$), $\Omega=5\un{rpm}$ ($\square$), $\Omega=6\un{rpm}$ ($\triangle$) and $\Omega=10\un{rpm}$ ($\diamond$) in semilogarithmic scale. The dash-dot line is given by $\mu =0.35 + 0.013 \log I$. (b): Variation of the ratio $k=\sigma_{xx}/\sigma_{zz}$ as a function of the inertial number $I$ for $\Omega=6\un{rpm}$. Inset: Variation of $k$ as a function of $I$ for for $\Omega=2\un{rpm}$, ($o$), $\Omega=4\un{rpm}$ ($\star$), $\Omega=5\un{rpm}$ ($\square$), $\Omega=6\un{rpm}$ ($\triangle$) and $\Omega=10\un{rpm}$ ($\diamond$) in semilogarithmic scale. The dash-dot horizontal line corresponds to $k=1$}
\label{rheology} 
\end{figure}

Typical variations of the effective friction coefficient $\mu$ as a function of the inertial number $I$ are plotted on Fig. \ref{rheology}a. A semilogarithmic representation (see inset of Fig. \ref{rheology}a) shows that data collected for different flow rates $Q$ collapse relatively well within the scaling:

\begin{equation} 
\mu = a +b\log I
\label{eqmuvsI}
\end{equation}
 
\noindent with $a\simeq0.35$ and $b\simeq0.013$ when $I$ ranges from $10^{-4}$ to $10^{-1}$. A departure from this scaling is observed when $I$ becomes smaller than $10^{-4}$. In this latter case, $\mu$ decreases more rapidly with $I$.
It is worth to note that the scaling given by Eq.\ref{eqmuvsI} is quantitatively similar to the one observed in the incline plane geometry \cite{Dacruz04_phd}, which suggests that both free surface flow and flows down to a rough incline plane may be described through the same constitutive laws. Relating $\mu$ and $I$ through a {\em local} constitutive law seems thus to be relevant. 

Figure \ref{rheology}b shows the variations of $k=\sigma_{xx}/\sigma_{zz}$ as a function of $I$. In the flowing layer \ie when $I$ exceed $I_{th}$, $k \rightarrow 1$. The {\em non monotonic} behaviour observed in the static phase is much more suprising: The parameter $k$ starts from a value {\em lower} than $1$ at the drum boundary $k(I\rightarrow 0) \simeq 0.8$, increases and reaches a maximum for $I \simeq 10^{-3}$ where $k(I \simeq 10^{-3}) \simeq 1.2$ and finally decreases for increasing $I$ and tends to $1$ within the flowing layer. Such observation is very different from the $k=1$ behaviour observed in the whole materials in both annular shear and incline geometry \cite{Silbert01_pre,Dacruz04_phd,Gdrmidi04_epje}.   

\begin{figure}[!tpb]
\centering
\includegraphics*[width=0.6\columnwidth]{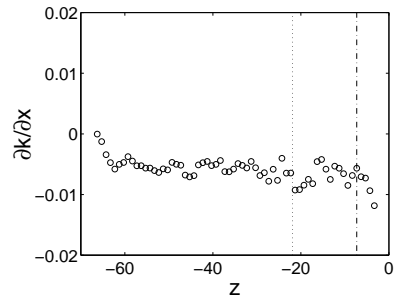}
\caption{Profiles of the $x$-gradient $\partial k/\partial x$ of the ratio $k = \sigma_{xx}/\sigma_{zz}$ at the center of the drum as observed for $\Omega=6\un{rpm}$. The vertical mixed lines and the vertical dotted lines show the free surface boundary and the static/flowing interface as defined in Fig.~\ref{comp6rpm} and Fig.~\ref{vit6rpm} respectively.}
\label{dkdx6rpm} 
\end{figure}

While the profile $\{\mu(z)\}$ is observed to be invariant along infinitesimal translation, the profile $\{k(z)\}$ is not (Fig. \ref{dkdx6rpm}).  The $x$-derivative of $k$ is found to be almost constant $\partial k/\partial x \simeq -0.05$ within the whole packing. In other words, the flow cannot be considered as homogeneous at the center of the drum as regard with the parameter $k$. Furthermore, while the curves $\mu(I)$ collected for different flow rates $Q$ collapse fairly well, the curves $k(I)$ do not. This strongly suggest that the non-local effects implied \eg by the existence of multi-scale rigid clusters embedded in the flow \cite{Bonamy02_prl} should be found in the constitutive law $k(I)$ rather than in $\mu(I)$ contrary to what was suggested in \cite{Bonamy02_prl,Bonamy03_epl,Bonamy03_gm,Gdrmidi04_epje}.

\section{Fluctuation analysis}

Let us now analyse the fluctuations $\delta v$ and $\delta \omega$ of the velocity and the vorticity respectively. Calling $\mathbf{c}(\mathbf{x},t)$  the "instantaneous" velocity of a bead located at the position $\mathbf{x}$ within the elementary slice $\Sigma$ at a given time $t$, the fluctuating part of the velocity $\delta \mathbf{c}(\mathbf{x},t)$ is defined as $\delta \mathbf{c}(\mathbf{x},t)=\mathbf{c}(\mathbf{x},t)-v_x(z)\mathbf{e}_x$ where $v_x(z)$ denotes the average velocity at the depth $z$ (see Fig.~\ref{vit6rpm}a). Profiles of velocity fluctuation $\delta v^2(z)$ are them computed by dividing $\Sigma$ into layers of one mean bead diameter wide, and averaging $\delta c^2$ over all the beads of the 400 frames whose center of mass is inside the corresponding layer. Same procedure is applied to determine the profiles of angular velocities fluctuations. In our athermal granular systems, the only time scale is provided by the velocity gradient $\partial_z v_x$. Therefore, we looked at profiles of $\{ \delta v/\partial_z v_x \}$ and $\{ \delta\omega/\partial_z v_x \}$ rather than direct profiles of $\{\delta v \}$ and $\{ \delta\omega \}$. 

\begin{figure}[!htpb]
\centering
\includegraphics*[height=0.6\columnwidth]{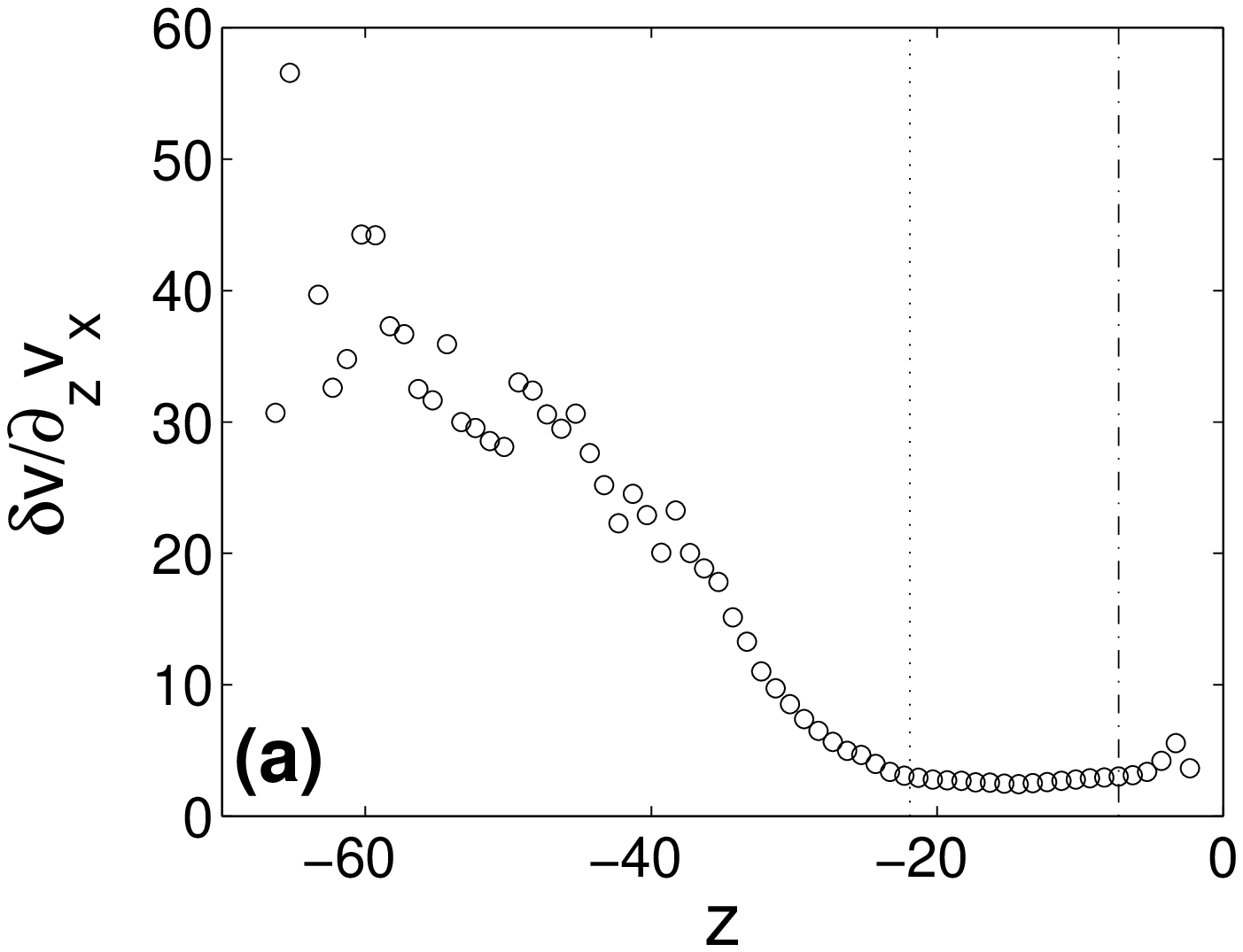}
\includegraphics*[height=0.6\columnwidth]{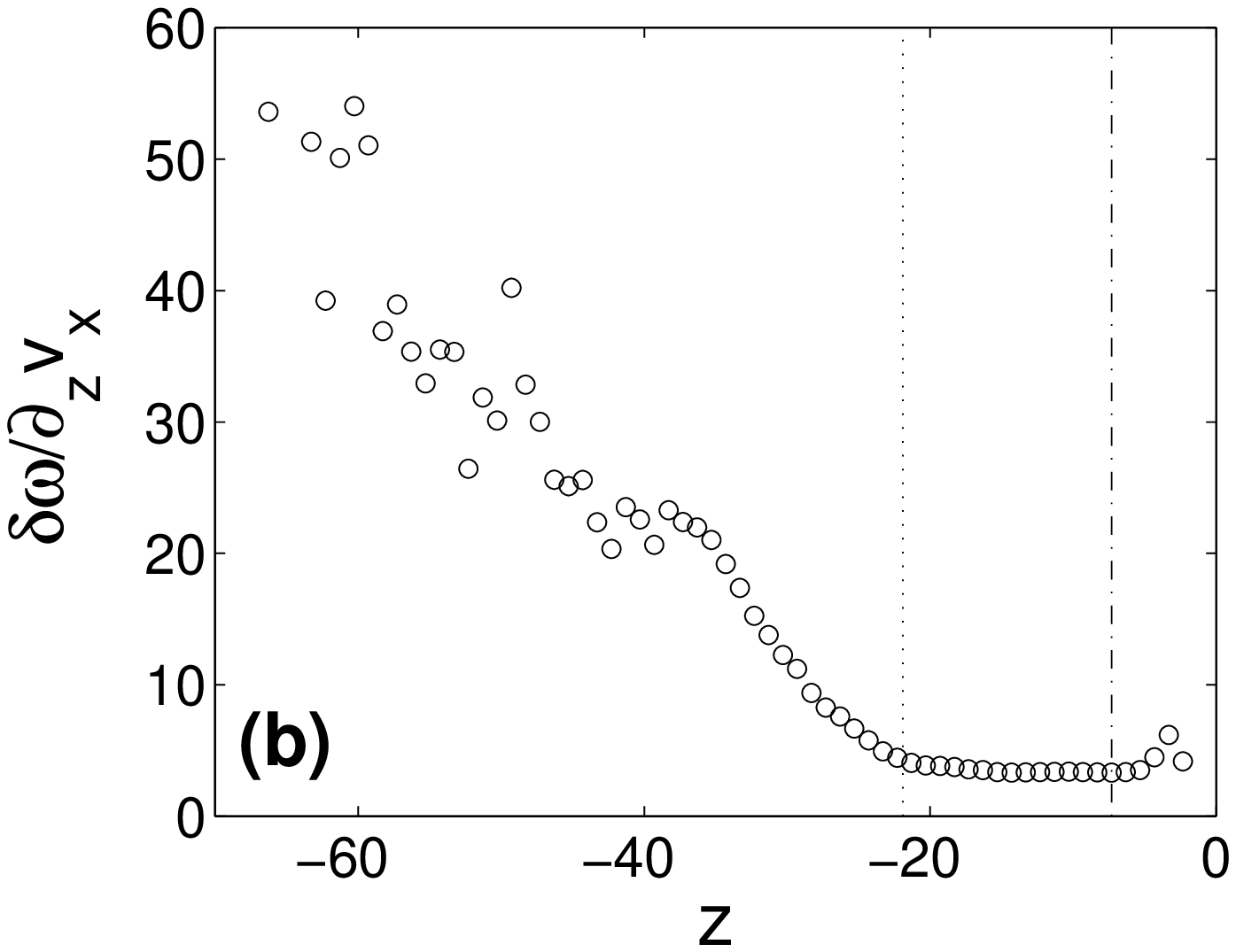}
\caption{Profile of translational velocity fluctuations $\delta v$ (a) and angular velocity fluctuation $\delta\omega$ (b) non-dimensionnalized by the shear rate $\partial_z v_x$ at the center of the drum obtained for $\Omega=6\un{rpm}$. The vertical mixed lines and the vertical dotted lines show the free surface boundary and the static/flowing interface as defined in Fig.~\ref{comp6rpm} and Fig.~\ref{vit6rpm} respectively.}
\label{fv6rpm}
\end{figure}

Figure \ref{fv6rpm} displays both translational velocity fluctuation profile (Fig. \ref{fv6rpm}a) and angular fluctuation (Fig. \ref{fv6rpm}b) nondimensionalized by the shear rate $\partial_z v_x$. In both cases, the nondimensionalized fluctuations are found to be constant within the flowing layer \ie:

\begin{equation}
\begin{array}{ll}
\frac{\delta v}{\partial_z v_x} \simeq 2.65 & \quad \mathrm{for} \quad z \geq -H \quad \mathrm{or} \quad I\geq I_{th}\\
\frac{\delta \omega}{\partial_z v_x} \simeq 3.35 & \quad \mathrm{for} \quad z \geq -H \quad \mathrm{or} \quad I\geq I_{th}
\end{array}
\label{fluct_flowinglayer}
\end{equation}

\begin{figure}[!htpb]
\centering
\includegraphics*[height=0.6\columnwidth]{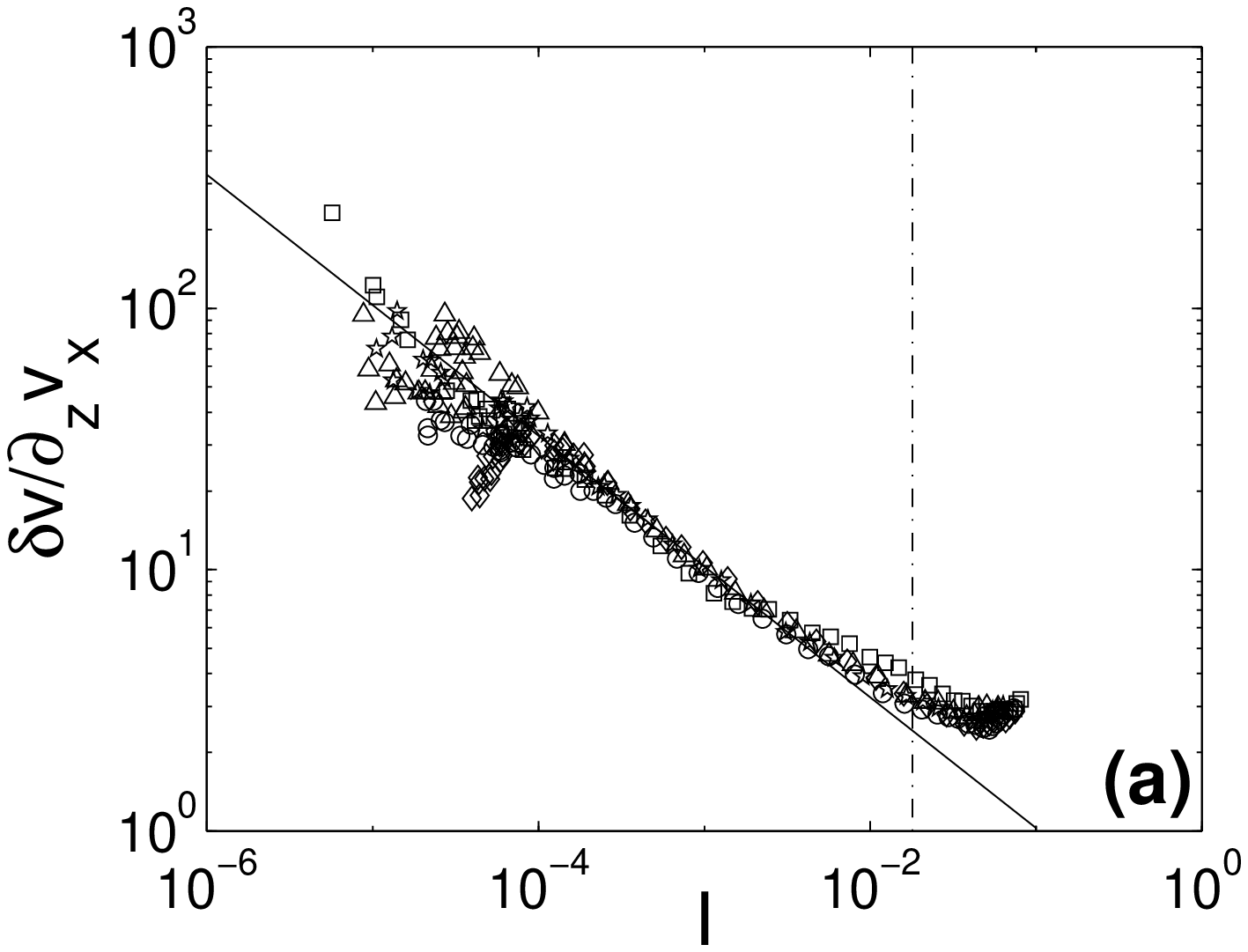}
\includegraphics*[height=0.6\columnwidth]{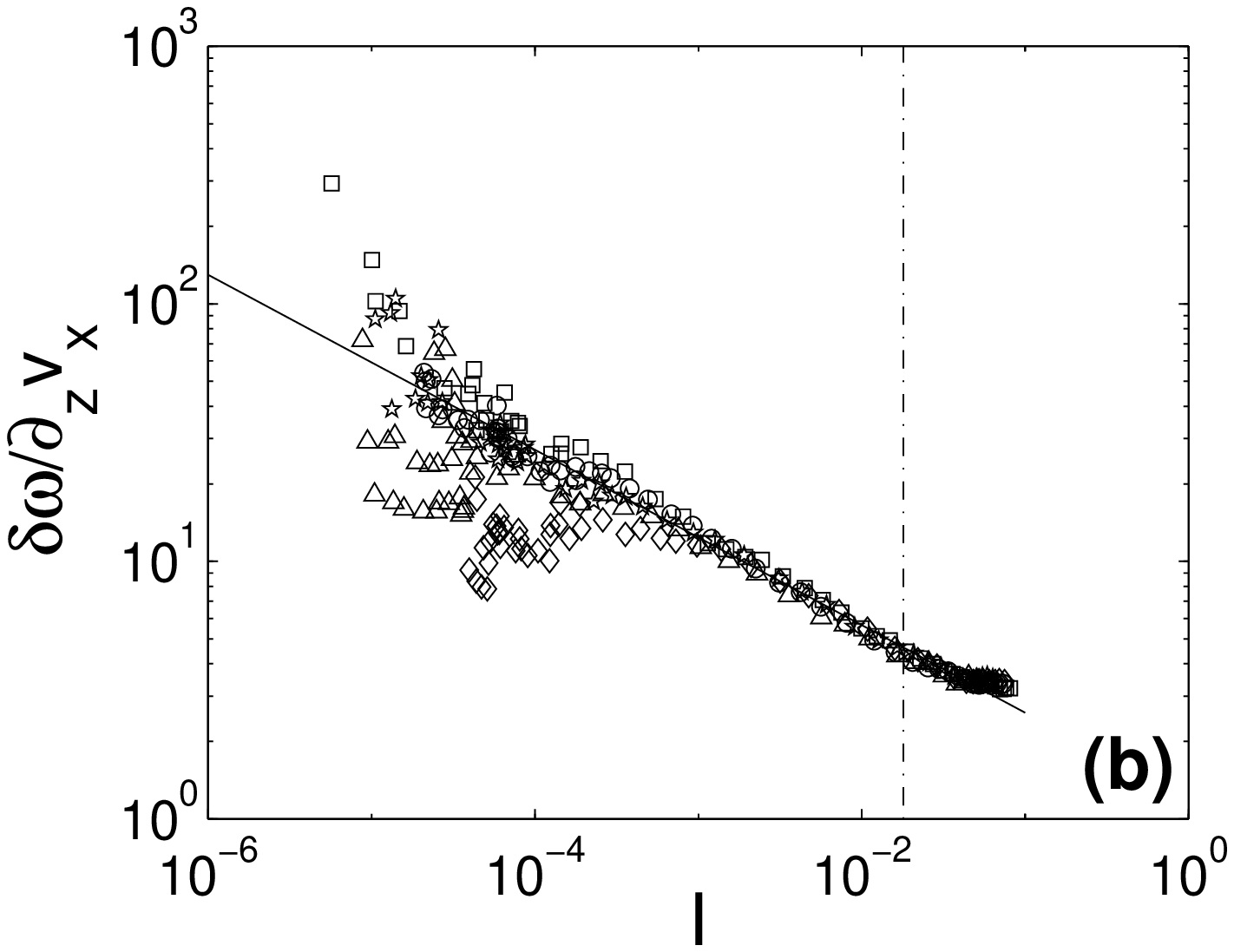}
\caption{(a) Non-dimensionalized velocity fluctuation $\delta v/\partial_z v_x$ as a function of the inertial number $I$ for $\Omega=2\un{rpm}$, ($o$), $\Omega=4\un{rpm}$ ($\star$), $\Omega=5\un{rpm}$ ($\square$), $\Omega=6\un{rpm}$ ($\triangle$) and $\Omega=10\un{rpm}$ ($\diamond$). The axes are logarithmic. The slope of the plain straight line is $-1/2$ (b) Non-dimensionalized velocity fluctuation $\delta\omega/\partial_z v_x$ as a function of the inertial number $I$ for $\Omega=2\un{rpm}$, ($o$), $\Omega=4\un{rpm}$ ($\star$), $\Omega=5\un{rpm}$ ($\square$), $\Omega=6\un{rpm}$ ($\triangle$) and $\Omega=10\un{rpm}$ ($\diamond$). The axes are logarithmic. The slope of the plain straight line is $-1/3$. In both graphes, the vertical dash-dot line show the static/flowing interface as defined by $I=I_{th}$.}
\label{fvvsI}

\end{figure}

\noindent In the static phase, both $\delta v/\partial_z v_x$ and $\delta\omega/\partial_z v_x$ are found to increase with the distance from the static/flowing interface. Figure \ref{fvvsI} plots both $\delta v / \partial_z v_x$ (Fig. \ref{fvvsI}a) and $\delta \omega / \partial_z v_x$ (Fig. \ref{fvvsI}b) as a function of the inertial number $I$. It evidences the existence of two different scalings within the static phase, namely:

\begin{equation}
\begin{array}{ll}
(a)\quad \frac{\delta v}{\partial_z v_x} \propto I^{-1/2} & \quad \mathrm{for} \quad I\leq I_{th}\\
(b)\quad \frac{\delta \omega}{\partial_z v_x} \propto I^{-1/3} & \quad \mathrm{for} \quad I \leq  I_{th}
\end{array}
\label{fluctvsI}
\end{equation}

Such scaling are very similar to the one observed in the shear geometry \cite{Dacruz04_phd}. The importance of these fluctuations with regards to the typical rate of deformation $\partial_z v_x$ (up to 40), as well as the scaling given by Eq.~\ref{fluctvsI}a exhibited within the static phase are compatible with the picture presented in \cite{Gdrmidi04_epje} to describe quasi-static flow: The average grains motion is made of a succession of very slow motions when the particle climbs over the next one, and a rapid motion when it is pushed back into the next hole by the confining picture.

\section{Concluding discussion}

Rheologies of 2D dense granular flows were investigated through {\em Non Smooth Contact Dynamics} simulations of steady surface flows in a rotating drum. Profiles of the different continuum quantities were measured at the center of the drum where the flow is non-accelerating. Volume fraction $\nu$ is found to be almost constant, around the Random Close Packing value $\nu^{RCP} \simeq 0.82$ within the whole packing, except for a tiny dilation (few percents) within the flowing layer, as expected from dilatancy effects. As observed experimentally \cite{Rajchenbach00_ap,Bonamy01_phd,Bonamy03_gm,Bonamy02_pof,Orpe01_pre,Jain02_pof,Felix02_phd}, the streamwise velocity profile $\{v_x(z)\}$ is found to be linear within the flowing layer, and to decrease exponentially within the static phase. Mean profile of the angular velocity was also measured at the center of the drum and was shown to be equal to half of the vorticity in the whole packing.

In a second step, profiles of the three independent component of the stress tensor were measured at the center of the drum. Quite surprisingly, the flow is found to be non-homogeneous at the center of the drum with regard to one of this component, namely $\sigma_{xx}$. In other words, $\partial_x \sigma_{xx}$ does not vanish whereas $\partial_x\nu$, $\partial_x {\bf v}$, $\partial_x \sigma_{zz}$ and $\partial_x \sigma_{xz}$ vanish.  

The inertial number $I$ - defined as the ratio between inertial solicitations and confinement solicitations was determined. This number is shown to be the relevant one to investigate quantitatively the rheology of the surface flows. The transition from the static to the flowing phase is found to occur to a fixed value $I_{th}$ of $I$,independently of the flow rate $Q$. Constitutive laws relating the components of the stress tensor to $I$ were determined. The effective friction$\mu=\sigma_{xz}/\sigma_{zz}$ is found to increase {\em logarithmically} with $I$, independently of the flow rate $Q$. This relation is found to match {\em quantitatively} the one observed in rough incline geometry. On the other hand, the ratio $k=\sigma_{xx}/\sigma_{zz}$ is found to be be {\em significantly different} from $k=1$ in contrast to what was observed in plane shear, annular shear, and rough incline geometry \cite{Silbert01_pre,Dacruz04_phd}. To be more precise, $k$ if found to vary non monotonically with $I$. Moreover, $\partial_x k$ is found not to vanish contrary to the x-derivative of the other continuum quantities. It is worth to note that $k=1$ together with a univocal relation between $\mu$ and $I$ would have naturally implied a Bagnold velocity profile, as observed in rough incline geometry, but not in the present free surface flow geometry. In other words, the selection of the velocity profile resides more in the function $k(I,...)$ than in $\mu(I)$.

Dependencies of $\{k(I)\}$ with Q, as well as the fact that $\partial_x k$ does not vanish lead us to conjecture that the ratio $k$ encodes the structure of the percolated network of grains in extended contact with each others  - referred to as the arches network. In this scenario, the structure of this network - and therefore the ratio $k$ - is related to the global geometry of the packing as well as to the orientation of the flow. This picture is broadly consistent with nonlocal models based on the coexistence of particle chains and fluidlike materials \cite{Mills99_epl,Bonamy03_epl}. However, a more detailed study is needed to verify this and understand how $k$ can be related to the global structure of the force network.

Finally, both velocity $\delta v$ and angular velocity $\delta \omega$ fluctuations were analysed. These quantities non-dimensionalized by the shear rate $\partial_z v_x$ were found to be constant - independent of the flow rate - within the flowing layer thickness. In the static phase, both $\delta v/\partial_z v_x$ and $\delta v/\partial_z v_x$ were found to decrease as different power-laws with $I$. This behaviour is consistent with the idea of an intermittent dynamics generated from the succession of rapid rearrangements and slow displacement \cite{Gdrmidi04_epje}. This change of behaviour at the static/flowing interface is broadly 
consistent with recent measurements of Orpe and Khakhar \cite{Orpe04_prl} revealing a sharp transition in the rms velocity distribution at this interface. Understanding what set precisely the scaling laws require precise statistical analysis of beads velocities at the grain scale. This represents interesting topic for a future investigation.
\begin{acknowledgments}
This work is supported by the CINES ({\em Centre Informatique National de l'Enseignement Sup\'erieur}), Montpellier-France, project mgc257.
\end{acknowledgments}


\begin{thebibliography}{10}
\bibitem{defAngle}
The procedure to define the angle between the $x$ axis and the horizontale is
  the following: (i) On each of the 400 snapshots of the sequences, one defines
  the free surface of the packing and calculate its slope; (ii) The angle
  $\theta$ is defined as the average of these slope over the 400 snapshots of
  the sequence.
\end{thebibliography}
\end{document}